\documentclass[fleqn,usenatbib]{mnras}

\usepackage{newtxtext,newtxmath}

\usepackage[T1]{fontenc}

\DeclareRobustCommand{\VAN}[3]{#2}
\let\VANthebibliography\thebibliography
\def\thebibliography{\DeclareRobustCommand{\VAN}[3]{##3}\VANthebibliography}

\usepackage{graphicx}	
\usepackage{amsmath}	
\usepackage{bm}

\newcommand{\Gaia}{\textit{Gaia} }
\newcommand{\Rdisc}{$R_\textrm{disc}$ }
\newcommand{\Rd}{$R_\textrm{d}$ }

\title[Disc Transition and Shrinking Post-GSE Merger]{Thick-to-thin disc transition and gas disc shrinking induced by the Gaia-Sausage-Enceladus merger}

\author[N. Funakoshi et al.]{Natsuki Funakoshi,$^{1}$\thanks{E-mail: \href{mailto:n.funakoshi@ucl.ac.uk}{n.funakoshi@ucl.ac.uk}}
Daisuke Kawata,$^{1}$
Jason L. Sanders,$^{2}$
Ioana Ciuc{\u{a}},$^{3}$
Robert J.~J. Grand,$^{4}$
\newauthor
HanYuan Zhāng$^{5}$
\\
$^{1}$Space and Climate Physics, Mullard Space Science Laboratory, University College London Holmbury St. Mary, Dorking, Surrey, RH5 6NT, UK\\
$^{2}$Department of Physics and Astronomy, University College London, London, WC1E 6BT, UK\\
$^{3}$Kavli Institute for Particle Astrophysics and Cosmology, Stanford University, 452 Lomita Mall, Stanford, CA 94305, UK\\
$^{4}$Astrophysics Research Institute, Liverpool John Moores University, 146 Brownlow Hill, Liverpool L3 5RF, UK\\
$^{5}$Institute of Astronomy, University of Cambridge, Madingley Road, Cambridge CB3 0HA, UK\\
}

\date{Accepted XXX. Received YYY; in original form ZZZ}

\pubyear{\the\year{}}

\begin{document}
\label{firstpage}
\pagerange{\pageref{firstpage}--\pageref{lastpage}}
\maketitle

\begin{abstract}
Understanding the Milky Way disc formation requires characterising its structural and kinematic properties as functions of stellar age. Using red giant stars from APOGEE DR17 and \Gaia DR3, we model the age-dependent stellar kinematics with a quasi-isothermal distribution function and fit disc parameters as a function of age using non-parametric splines. We identify a transition from thick to thin disc populations around 10~Gyr ago. Stars older than this have short scale lengths ($\sim$1.7~kpc), typical of the thick disc, while younger stars exhibit increasing scale length with decreasing age, consistent with inside-out formation of the thin disc. This transition possibly coincides with the end of the starburst triggered by the Gaia-Sausage-Enceladus (GSE) merger. Stars formed around 10~Gyr ago exhibit a dip in scale length, even shorter than that of the thick disc. Comparison with an Auriga simulation suggests that this scale-length dip reflects gas disc shrinking caused by the transition from a cold to hot gas accretion mode. We propose the following disc formation scenario: (1) the thick disc formed under cold-mode accretion; (2) the GSE merger triggered a starburst and increased the total mass of the Galaxy, causing the transition to hot-mode accretion; (3) rapid gas consumption led to temporary shrinking of the star-forming gas disc; and then (4) thin disc grows in an inside-out fashion, as the size of the star-forming gas disc grows via hot-mode smooth gas accretion.
\end{abstract}

\begin{keywords}
Galaxy: disc -- Galaxy: evolution -- Galaxy: kinematics and dynamics
\end{keywords}



\section{Introduction}\label{sec:intro}

Gaia-Sausage-Enceladus (GSE) is believed to be the last significant merger event experienced by the Milky Way.
The GSE remnants are identified through the kinematics of metal-poor halo stars typically with low-[$\alpha$/Fe] abundances \citep{2018MNRAS.478..611B,2018Natur.563...85H,2018ApJ...863..113H}.
The merger is believed to have occurred between 8 and 11~Gyr ago \citep[e.g.,][]{2020ARA&A..58..205H,2019MNRAS.482.3426M,2019A&A...632A...4D,2019NatAs...3..932G,2020MNRAS.496.1922B,2021NatAs...5..640M}
and had a significant impact on the Galactic disc \citep[e.g.,][]{2020MNRAS.496.1922B,2020MNRAS.497.1603G,2021MNRAS.503.5846R}. 

Observations show that the Galactic disc consists of two distinct components \citep{1982PASJ...34..365Y,1983MNRAS.202.1025G}, although there is ongoing debate as to whether the thin and thick discs should be regarded as truly distinct components \citep[e.g.,][]{2007ApJ...663L..13B,2012ApJ...751..131B,2016AN....337..976K,2017A&A...608L...1H}. The first is the high-[$\alpha$/Fe] disc, which is older and geometrically thicker—hence referred to as the "thick disc". The second is the younger low-[$\alpha$/Fe] disc, which is typically geometrically thin and known as the "thin disc"\footnote{Flaring in the outer disc means the low-[$\alpha$/Fe] population can be geometrically thick in some parts of the Galaxy \citep[e.g.,][for reviews]{2016AN....337..976K,2024arXiv241212252K}. However, for convenience, we refer to the chemically defined older high-[$\alpha$/Fe] disc and younger low-[$\alpha$/Fe] disc as the thick and thin discs, respectively.} \citep[e.g.,][]{1998A&A...338..161F,2000AJ....120.2513P,2003A&A...397L...1F,2014A&A...562A..71B,2015ApJ...808..132H}. 
Recent years have seen theories develop that connect together the GSE merger with the configuration of the Galactic discs. By analysing a series of high-resolution cosmological magnetohydrodynamical simulations \citep[\texttt{Auriga:}][]{2017MNRAS.467..179G,2024MNRAS.532.1814G}, \citet{2020MNRAS.497.1603G} showed that a gas-rich merger can trigger a central starburst, leading to a transition from thick to thin disc formation \citep[see also][]{2004ApJ...612..894B,2012MNRAS.426..690B}. Motivated by this result, it has been proposed that the GSE merger may have caused a similar transition in the Milky Way.

This theoretical scenario finds support in observational results. \citet{2024MNRAS.528L.122C} analysed red giant stars from APOGEE (The Apache Point Observatory Galactic Evolution Experiment), whose ages were estimated using BINGO (Bayesian INference for Galactic archaeOlogy), a Bayesian machine learning framework developed by \citet{2021MNRAS.503.2814C}. 
Their analysis revealed a distinct phase of rapid increase of [Fe/H] and decrease of [$\alpha$/Fe] in stars aged between 12 and 10~Gyr \citep[on the \emph{relative} age scale used in][]{2021MNRAS.503.2814C}. 
They interpreted this as evidence for a major starburst episode, which they named the Great Galactic Starburst (GGS). This event was likely driven by a rapid gas inflow that began around 12~Gyr ago and continued until roughly 10~Gyr ago.
Comparisons with Auriga simulations suggest that the GGS was likely driven by the gas-rich GSE merger event.
\citet{2024MNRAS.528L.122C} indicated that after this gas-rich merger event, the high-[Fe/H] and low-[$\alpha$/Fe] thin disc population began to form. This suggests that the transition in the disc formation phase—from thick to thin—occurred after the gas-rich GSE merger.

A key question is whether \textit{structural and kinematical} properties of the disc also exhibit a clear transition from the thick to the thin disc around the same epoch as GSE merger. 
Many studies have analysed the age-dependent structure of the Galactic disc to understand how the disc has evolved.
For example, \citet{2019ApJ...884...99F} modelled the ages, metallicities and radial distribution of APOGEE red clump stars and found clear evidence for inside-out growth of the low-[$\alpha$/Fe] thin disc.
\citet{2023NatAs...7..951L} estimated the surface brightness profile of the Milky Way as a function of stellar age. They revealed a broken radial structure, indicating a more complex and extended disc than previously assumed based on single exponential models \citep[e.g., see also][]{2017MNRAS.471.3057M, 2022MNRAS.513.4130L}. \citet{2025A&A...700A..89K} used APOGEE DR17 data and an orbit superposition modelling. They identified two distinct age-metallicity sequences in the Milky Way disc, which they associated with the formation histories of an older inner disc and a younger outer disc.
\citet{2025ApJ...990..203I} demonstrated that the disc scale length, scale height and flaring are strongly correlated with stellar age, highlighting the importance of time-dependent processes in the formation and evolution of the Milky Way disc.

\citet{2023MNRAS.521.1462Z} analysed the velocity distribution of the stellar disc as a function of stellar age
to calibrate the period-age relation for O-rich Mira variable stars.
They estimated the structural properties of the disc by fitting the velocity distributions of individual stars with an action-based distribution function (hereafter, DF) after dividing their sample into different age bins.
This method does not require knowledge of the stellar density distribution and is less affected by spatial selection bias in observational data, since the kinematics of stars are unlikely to depend on the observational selection function.

Following the approach of \citet{2023MNRAS.521.1462Z}, we investigate the kinematical distribution of red giant stars in APOGEE DR17 as a function of age. We use stellar ages measured by \citet{2024MNRAS.528L.122C} and stellar kinematics from APOGEE and \Gaia DR3. 
By fitting action-based DFs as a function of age, for the first time we successfully trace the age dependence of the structural and kinematic properties of the Galactic disc seamlessly over the full stellar age range.
Particular attention is paid in the age dependence of the scale length of the radial surface density profile of the Galactic disc. \citet{2018MNRAS.474.3629G} suggested that the transition from thick to thin disc formation phase involved a shrinking of the star-forming gas disc analysing an \texttt{Auriga} cosmological simulation. They showed that the transition of the gas accretion mode contributes to the thick-to-thin disc formation phase transition and this mode transition causes a temporary gas supply gap, which leads to the gas shortage and shrinking of the star-forming gas disc. If this is the case, the scale length of the stars formed in this transition phase would be smaller in the scale length, compared to the other period.

In Section~\ref{sec:data}, we describe the dataset of red giant stars used in this study. Section~\ref{sec:modelling} outlines our modelling method based on an action-based quasi-isothermal DF. 
Section~\ref{sec:result} presents our results of fitting the observational data with the DFs as a function of age. Our discussion in Section~\ref{sec:discussion} is divided into three parts: in Section~\ref{sec:kinematical_footprint}, we analyse which physical properties in the data drive the trends found in the DF fitting results; in Section~\ref{sec:auriga}, we apply the same DF fitting method to Auriga simulation data and compare the resulting trends with those observed in the Milky Way; in Section~\ref{sec:indication}, we discuss the implications of our results in the context of the evolution of the Galaxy. Section~\ref{sec:summary} provides a summary of this study.

\section{Data}\label{sec:data}
\begin{figure}	\includegraphics[width=\columnwidth]{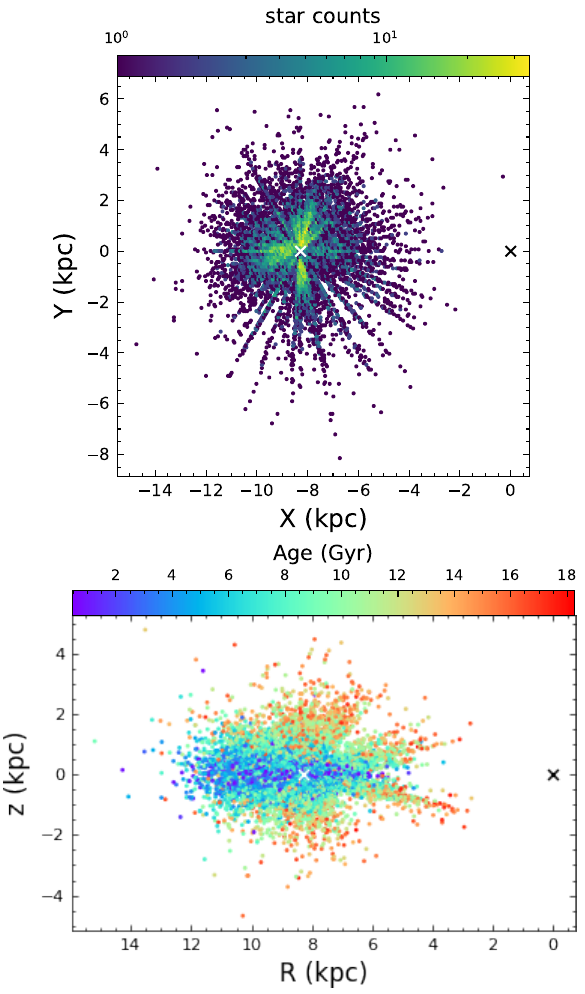}
    \caption{Spatial distribution of the red giant stars used in our study. The upper panel shows the face-on view of the stellar density distribution coloured with the logarithmic density. The lower panel presents the distribution in Galactocentric radius and vertical height, where colour denotes stellar age. The position of the Sun is indicated by a white cross in both panels, located at ($X,Y$)=($-8.275$,0)~kpc in the upper panel and ($R,z$)= (8.275,0.0208)~kpc in the lower panel. The Galactic centre is marked by a black cross.}
    \label{fig:xyrz}
\end{figure}

The red giant sample was taken from \citet{2024MNRAS.528L.122C}.
\citet{2021MNRAS.503.2814C} developed BINGO, a supervised Bayesian Neural Network 
to estimate stellar ages and their uncertainties of red giant stars using APOGEE-2 DR17 \citep{2015AJ....150..173N,2017AJ....154...94M,2019PASP..131e5001W}. The input features are stellar parameters $T_\textrm{eff}$, $\log g$ and [Fe/H], and chemical abundance ratios, [C/Fe], [N/Fe] and [Mg/Fe], which were derived with APOGEE Stellar Parameters and Chemical Abundances Pipeline \citep[ASPCAP;][]{2016AJ....151..144G} using the line lists described in \citet{2015ApJS..221...24S} and \citet{2021AJ....161..254S}. Its training data were sourced from the APOKASC-2 stars with asteroseismic ages calculated by \citet{2021A&A...645A..85M}.
Since asteroseismic ages are more reliable for stars on the Red Giant Branch (RGB) and high-mass (>1.8 $\mathrm{M_\odot}$) Red Clump (RC) stars, the training data were restricted to these populations. 
In addition, \citet{2024MNRAS.528L.122C} applied quality cuts of \texttt{ASPCAPFLAG=0}, signal-to-noise ratio$>100$, $1 < \log g < 3.5$ and $4000 \mathrm{~K}< T_\textrm{eff} < 5500\mathrm{~K}$. Further, they developed a classifier to select only the RGB and the high-mass RC stars, and applied it to the APOGEE data to select the stars similar to the training data. After applying the trained BINGO model to these stars, they estimated ages for 89,591 stars with their statistical uncertainties. These estimated ages are independently validated by their consistency with spectroscopic ages of APOGEE red giant stars derived using \texttt{XGBoost} \citep{2016arXiv160302754C} model trained on asteroseismic data \citep[see Fig. 6 of][]{2023A&A...678A.158A}.
It should be noted that some stars in the sample have estimated ages older than the age of the Universe. This arises from the non-informative age prior used by \citet{2021A&A...645A..85M}, which has a high upper limit of 40~Gyr. Nevertheless, \citet{2021MNRAS.503.2814C} discussed that the age estimates are reliable in terms of relative age. Hence, throughout this paper, we adopt the specific age scale used in BINGO, which may not correspond to the absolute age scale of the Universe. 

We cross-matched the APOGEE sample with the \Gaia DR3 sources \citep{2023A&A...674A..37G}, to obtain proper motions and parallaxes. $V_{\mathrm{helio}}$ from the APOGEE catalogue was adopted as the radial velocity. We applied quality cuts on the \Gaia data by selecting the data with $\texttt{parallax}/\texttt{parallax\_error}>5$ and $\texttt{RUWE}<1.4$ \citep{2021A&A...649A...2L}. We only included stars with an uncertainty of less than 0.02 dex in the base-10 logarithmic stellar age estimate and a probability higher than 95\% of being an RGB or high-mass RC star. 
Note that the age uncertainty of BINGO represents aleatoric uncertainty of the neural network model predictions. This uncertainty reflects how well the Neural Network model can replicate the training data,
and can be smaller than the measured uncertainties of the training data \citep{2021MNRAS.503.2814C}. 
The age uncertainty of BINGO can still be used to select the stars with high-confidence age inference.

In addition, we removed stars with vertical distances greater than $5$ times the dispersion of the vertical distribution of stars with similar ages. This process removed three stars with stellar ages of less than 1~Gyr. In total, 16,617 star samples were obtained. 
Most of the stars in our sample are disc stars with [Fe/H]$>-1.0$. This is because stars with lower metallicity ([Fe/H]$<-1.0$) have larger age uncertainties of BINGO. The training data include very few such low-metallicity stars due to the \textit{Kepler} astroseismic survey's narrow field of view, which is restricted to the Galactic disc. With so few examples in training data, the neural network cannot reliably constrain the ages for the low-metallicity stars. 
As a result, such stars rarely meet our age-uncertainty requirement ($<0.02$ dex). 
This selection effect is, however, appropriate for our purposes, because it naturally reflects the training data, which mainly consist of Galactic disc stars, and keeps the sample focused on the disc population.

The spatial distribution of our selected data is shown in Fig.~\ref{fig:xyrz}. Observational data are sparse within 5~kpc of the Galactic centre. Since the effects of the Galactic bar and bulge are likely to be prominent within this inner region, it is reasonable to exclude the bar-dominated area from the analysis when using an axisymmetric model, as adopted in this study (see Section~\ref{sec:modelling}).

\section{Modelling}\label{sec:modelling}
We aimed to fit the stellar kinematics data with a probability DF of stars, $p(\bm{\mu}, v_{\|} \mid \ell, b, \varpi, \log_{10}\tau)$, where $\bm{\mu}$ and $ v_{\|}$ are the proper motion vector and the line-of-sight velocity, respectively, $\ell$ and $b$ are the Galactic longitude and latitude and $\tau$ is the stellar age. $\varpi$ is the \Gaia DR3 parallax, which was corrected for the zero-point offset based on \citet{2021A&A...649A...4L} using the Python package \texttt{gaiadr3\_zeropoint}\footnote{\url{https://pypi.org/project/gaiadr3-zeropoint/}}. Since stellar density measurements are highly sensitive to observational selection functions and are difficult to obtain accurately, we focused on fitting (action-based) DFs using only stellar velocities, following \citet{2023MNRAS.521.1462Z} and \citet{2024MNRAS.530.2972S}.

We adopted \( R_0 = 8.275 \, \mathrm{kpc} \) \citep{2021A&A...647A..59G} as the Galactocentric radius of the Sun, and \( z_0 = 20.8 \, \mathrm{pc} \) \citep{2019MNRAS.482.1417B} as its vertical height above the Galactic mid-plane. For the solar motion, we used \( (u_0, v_0, w_0) = (9.65, 14, 8.59) \, \mathrm{km~s^{-1}} \) \citep{2024MNRAS.529.1035A}, where $u_0, v_0$ and $w_0$ are the components of the solar proper motion with respect to the Local Standard of the Rest in the direction of the Galactic centre, the Galactic rotation and the north Galactic pole, respectively. The circular velocity at the solar radius, $V_\mathrm{c}(R_0)$, was taken from a fixed axisymmetric gravitational potential of the Milky Way described in \citet{2017JOSS....2..388P} and was set to \( 231.21 \, \mathrm{km~s^{-1}} \).

We begin by writing the probability DF as follows:
\begin{equation}\label{eq:probabiliry}
p(\bm{\mu}, v_{\|} \mid \ell, b, \varpi, \log_{10}\tau)=\frac{p(\ell, b, \varpi, \bm{\mu}, v_{\|}, \log_{10}\tau)}{p(\ell, b, \varpi, \log_{10}\tau)},
\end{equation}
where both the numerator and denominator have a cancelling contribution from the selection function.
We accounted for uncertainties in the proper motion, line-of-sight velocity, parallax and stellar age by marginalizing over them.  The numerator is evaluated as follows;
\begin{equation}\label{eq:numeoflikelihood}
\begin{aligned}
&p\left(\ell, b, \varpi, \bm{\mu}, v_{\|}, \log_{10}\tau\right) \\
=& \int \mathrm{d}^2 \bm{\mu}^{\prime} \mathrm{d} v_{\|}^{\prime} \mathrm{d} \varpi^{\prime} \mathrm{d} \log_{10}{\tau}^{\prime} \mathcal{N}\left(\bm{\mu}^{\prime} \mid \bm{\mu}, \mathbf{\Sigma}_\mu\right) \mathcal{N}\left(\varpi^{\prime} \mid \varpi, \sigma_\varpi^2\right) \\
&\mathcal{N}\left( v_{\|}^{\prime} \mid v_{\|}, \sigma_{v_{\|}}^2\right) \mathcal{N}\left( \log_{10}{\tau}^{\prime} \mid \log_{10}{\tau}, \sigma_{\log_{10}\tau}^2 \right)
p\left(\ell, b, \varpi^{\prime}, \bm{\mu}^{\prime}, v_{\|}^{\prime}, \log_{10}{\tau}^{\prime} \right) ,
\end{aligned}
\end{equation}
where $\mathcal{N}\left(x\mid\mu,\sigma^2\right)$ denotes a Gaussian with mean $\mu$ and variance $\sigma^2$. The sky position of stars, $(\ell,b)$, was assumed to be accurately measured and was not included in the marginalisation.
Then, we related the DF of the observable coordinates to a DF of action as follows:
\begin{equation}\label{eq:propto_s5}
p\left(\ell, b, \varpi^{\prime}, \bm{\mu}^{\prime}, v_{\|}^{\prime}, \log_{10}\tau\right)=\left|\frac{\partial(\bm{J}, \bm{\theta})}{\partial\left(\ell, b, \varpi, \bm{\mu}, v_{\|}\right)}\right| f_{\tau}(\bm{J}) \propto s^6 \cos b \text { }f_{\tau}(\bm{J}) \text {, }
\end{equation}
where $\bm{J}=(J_R,J_\phi,J_z)$ is the vector of actions corresponding to the observable coordinates in six dimensions (with the corresponding angle coordinates, $\bm{\theta}$) and $s$ is the distance corresponding to the parallax. The factor $s^6\cos{b}$ is derived from the Jacobian between the Galactic coordinates and Cartesian coordinates, $(\bm{x},\bm{v})$. The Jacobian between $(\bm{x},\bm{v})$ and $(\bm{J},\bm{\theta})$ is unity due to the canonical transformation.
The factor $s^4\cos{b}$ appears in the transformation of a volume element in three-dimensional position space, which includes $s=1/\varpi$, while $s^2$ appears in the transformation of an area element in proper motion velocity space. Here, the subscript $\tau$ in $f_\tau(\bm{J})$ means that the parameters of the action-based DF are functions of stellar age $\tau$, so that the DF varies with stellar age $\tau$.

Adopting a quasi-isothermal DF, $f(\bm{J})$, from \citet{2010MNRAS.401.2318B}, we used an implementation provided in \texttt{AGAMA} \citep{2018arXiv180208255V}, which takes the following form:
\begin{equation}\label{eq:DF}
\begin{gathered}
f_{\tau}(\bm{J})=\frac{\tilde{\Sigma} \Omega}{2 \varpi^2 \kappa^2} \times \frac{\kappa}{\tilde{\sigma}_r^2} \exp \left(-\frac{\kappa J_R}{\tilde{\sigma}_r^2}\right) \times \frac{\nu}{\tilde{\sigma}_z^2} \exp \left(-\frac{\nu J_z}{\tilde{\sigma}_z^2}\right) \times B\left(J_\phi\right), \\
B\left(J_\phi\right)= \begin{cases}1 & \text { if } J_\phi \geq 0, \\
\exp \left(\frac{2 \Omega J_\phi}{\tilde{\sigma}_r^2}\right) & \text { if } J_\phi<0,\end{cases} \\
\tilde{\Sigma}\left(R_{\mathrm{c}}\right) \equiv \Sigma_0 \exp \left(-R_{\mathrm{c}} / R_{\mathrm {disc }}(\tau)\right), \\
\tilde{\sigma}_r^2\left(R_{\mathrm{c}}\right) \equiv \sigma_{R, 0}^2(\tau) \exp \left(-2\left(R_{\mathrm{c}}-R_0\right) / R_{\sigma, R}(\tau)\right), \\
\tilde{\sigma}_z^2\left(R_c\right) \equiv \sigma_{z, 0}^2(\tau) \exp \left(-2\left(R_c-R_0\right) / R_{\sigma, z}(\tau)\right),
\end{gathered}
\end{equation}
where $R_c$ is the guiding radius, defined as the radius of a circular orbit with angular momentum, $J_{\phi} \equiv L_\mathrm{z} = R_{\textrm{c}}\times V_\mathrm{c}(R_\textrm{c})$. Note that $L_\mathrm{z}$ is a conserved quantity in axisymmetric potentials and can also be expressed as $L_\mathrm{z}=R\times V_{\phi}$ using the star's current Galactocentric radius and azimuthal velocity. The quantities $\kappa,\Omega$ and $\nu$ denote the epicyclic, angular and vertical frequencies of the orbit, respectively. The DF is described with five parameters: $R_\mathrm{disc}$, $\sigma_{R,0}$, $\sigma_{z,0}$, $R_{\sigma,R}$ and $R_{\sigma,z}$, where \Rdisc is the radial scale length, $\sigma_{R,0}$ and $\sigma_{z,0}$ are the normalisation parameters for the radial and vertical velocity dispersions of the stars at $R_c=R_0$ and $R_{\sigma,R}$ and $R_{\sigma,z}$ represent the radial scale lengths of the radial and vertical velocity dispersion profiles, respectively. Table~\ref{tab:params_sum} summarises these parameters along with the priors described in the following subsections. It is worth noting that these parameters describe the shape of the DF and approximate the behaviour of physical quantities, but they are not strictly equivalent to them \citep[see also][]{2011MNRAS.413.1889B,2013ApJ...779..115B}. For example, the radial scale length parameter, $R_\mathrm{disc}$, controls the exponential fall-off of the DF in guiding radius, but the actual stellar density profile is also influenced by the shape of the gravitational potential. Nevertheless, the resulting stellar density profile is still approximately exponential, with a scale length similar to $R_\mathrm{disc}$ \citep[e.g., Fig.~6 of][]{2010MNRAS.401.2318B}.
Although previous studies have shown that the radial surface density profile for mono-age disc population is better described by a broken exponential, especially for the low-[$\alpha$/Fe] disc \citep[e.g.,][]{2017MNRAS.471.3057M,2024NatAs...8.1302L}, here we model the disc with a single exponential scale length. This enables us to trace the stellar age dependence of the overall structure of the disc with fewer parameters.

We adopted a fixed axisymmetric gravitational potential of the Milky Way from \citet{2017JOSS....2..388P} and computed the actions, $\bm{J}=\bm{J}(\bm{x},\bm{v})$, of each star using the St\"ackel fudge approximation as implemented in \texttt{AGAMA}.

\subsection{Importance sampling}
The denominator in equation~\eqref{eq:probabiliry}, $p(\ell,b,\varpi,\log_{10}\tau)$, involves the Monte Carlo integrals over the full three-dimensional velocity space, making it computationally expensive. 
To compute the denominator efficiently, we applied importance sampling, as described by \citet{2023MNRAS.521.1462Z} and \citet{2024MNRAS.530.2972S}, 
Here, a sample of the three-dimensional velocities, $\bm{v}$, used for integration was drawn from a probability DF, $G_{\tau}(\bm{v} \mid \ell, b, \varpi)$. This function is proportional to an assumed quasi-isothermal DF, $f_\tau^{\prime}(\bm{J})$, with fixed parameters. 
Importance sampling integration is more accurate when the sampling function, $f^{\prime}(\bm{J})$, closely resembles that of the function being integrated, $f(\bm{J})$. As explained in Section~\ref{sec:result}, we found that the parameters of our DF that best describe the observational data vary with stellar age.
and are expressed as an age-dependent function $f_\tau(\bm{J})$. 
Therefore, we used an age-dependent sampling function of $f_\tau^{\prime}(\bm{J})$.
Given a stellar age $\tau$ and a 3D position $(\ell, b, \varpi)$, the probability DF can be drawn as follows:
\begin{equation}\label{eq:G}
G_{\tau}(\bm{v} \mid \ell, b, \varpi)=\frac{p(\ell, b, \varpi, \bm{v}, \log_{10}\tau)}{\int \mathrm{d}^3 \bm{v}\text{ } p(\ell, b, \varpi, \bm{v}, \log_{10}\tau)} = \frac{f^{\prime}_\tau(\bm{J})}{\int \mathrm{d}^3 \bm{v} \text{ } f^{\prime}_\tau(\bm{J})}\text{. }
\end{equation}
To minimise the bias in the Monte Carlo integration, we adjusted the parameters of $f^{\prime}_\tau(\bm{J})$ so that it approximates the true distribution $f_\tau(\bm{J})$.
This adjustment was done iteratively by repeating the entire Markov Chain Monte Carlo (MCMC) fitting process, each time updating the parameters of $f^{\prime}_\tau(\bm{J})$ based on the results of the previous run.

The denominator on the right side of equation~\eqref{eq:G} can be computed using \texttt{AGAMA}. We generate a set of $N=1280$ samples for each star using the MCMC performed with \texttt{emcee} \citep[][]{2013PASP..125..306F}, sampling velocities $\bm{v}=(v_x,v_y,v_z)$ based on $G_{\tau_k}(\bm{v} \mid \ell_k, b_k, \varpi_k)$ as the likelihood for the $k$-th star at $(\ell_k, b_k, \varpi_k)$ and $\tau_k$.
Also, $N=1280$ random parallaxes and stellar ages are drawn from Gaussian distributions, $\mathcal{N}(\varpi_k,\sigma^{2}_{\varpi_k})$ and $\mathcal{N}(\log_{10}\tau_k,\sigma^{2}_{\log_{10}{\tau_k}})$, respectively, to account for the uncertainty in parallax and stellar age.
The integral of the denominator in equation~\eqref{eq:probabiliry} for each star is then computed as follows,
\begin{align}\label{eq:important-deno}
p(\ell_k, b_k, &\varpi_k, \log_{10}\tau_k)
\approx \frac{A}{N} \times \notag \\
&\quad \sum_i^{\substack{\varpi_i\text{ from } \mathcal{N}(\varpi_k,\sigma_{\varpi_k}^2)  \\
\bm{v}_i \text{ from } G(\bm{v} \mid \ldots) \\
\tau_i \text{ from } \mathcal{N}(\log_{10}{\tau_k},\sigma^{2}_{\log_{10}{\tau_k}})}}
s_i^4 \cos b \,
\frac{f_{\tau_i}\left(\bm{J}\left( \ell_k,b_k,\varpi_i,\bm{v}_i\right)\right)}
{f_{\tau_k}^{\prime}\left(\bm{J}\left( \ell_k,b_k,\varpi_k,\bm{v}_i\right)\right)}
\text{,}
\end{align}
where
\begin{align}\label{eq:A}
A&=\int \mathrm{d}^3 \bm{v} \text{ } f_{\tau_k}^{\prime}(\bm{J}(\ell_k,b_k,\varpi_k,\bm{v}))\nonumber\\
&=\sum_i^{\bm{v}_i \text{ from } G(\bm{v} \mid \ldots)}f_{\tau_k}^{\prime}(\bm{J}(\ell_k,b_k,\varpi_k,\bm{v}_i)).
\end{align}
Here, $k$ indexes stars and $i$ indexes samples per star.
The term $s^4$ $\cos{b}$ represents the Jacobian for transforming a Cartesian spatial volume into Galactic coordinates.
The denominator of equation~\eqref{eq:important-deno}, $f_{\tau_k}^{\prime}(\bm{J})$ is precomputed once since the parameters of $f_{\tau_k}^{\prime}$ is fixed during the MCMC fitting process.
The action variables $\bm{J}$ used here are computed for each star using the fixed spatial coordinates ($\ell_k, b_k, \varpi_k$) and a velocity component $\bm{v}_i$ drawn from $G_{\tau_k}(\bm{v} \mid \ell_k, b_k, \varpi_k)$.  
On the other hand, for the numerator of equation~\eqref{eq:important-deno}, since the age dependence of the parameters for $f_{\tau_i}$ keeps being updated at each step of MCMC, $f_{\tau_i}(\bm{J})$ needs to be computed at every step of MCMC fitting.
$\bm{J}$ for each star is computed using sampled $\varpi_i, \bm{v}_i$ and $\tau_i$.
The constant factor $A$ defined in equation~\eqref{eq:A} is precomputed once for each star in the same manner as the denominator of equation~\eqref{eq:important-deno}. 
Nevertheless, thanks to importance sampling, the computational cost of three-dimensional integration is effectively reduced to a simple summation, as shown in equation~\eqref{eq:important-deno}.

\subsection{Likelihood and priors}
For fitting the data, we maximise the following log-likelihood using MCMC,
\begin{equation}\label{eq:loglikelihood}
\ln L=\sum_k^{\text {stars }} \ln p\left(\bm{\mu}_k, v_{\|k} \mid \ell_k, b_k, \varpi_k, \log_{10}\tau_k\right),
\end{equation}
for the five parameters: $R_{\textrm{disc}}$, $\sigma_{R,0}$, $\sigma_{z,0}$, $R_{\sigma,R}$, $R_{\sigma,z}$. Since our primary interest is to investigate how these parameters vary with stellar age, we model them as functions of age, $\tau$ (Gyr), i.e., $R_\mathrm{disc}(\tau),\sigma_{R,0}(\tau),\sigma_{z,0}(\tau),R_{\sigma,R}(\tau)$ and $R_{\sigma,z}(\tau)$. Their age dependence is modelled using flexible cubic spline functions, following \citet{2024MNRAS.530.2972S}. We found that 10 knots provides a good balance between capturing the age variation of these parameters and avoiding overfitting. These knots were evenly spaced across the age range between 0.46 to 18.26~Gyr, which are the minimum and maximum stellar ages in our observational data.
The five parameters for the DFs are adjusted at these 10 knots to fit the data, resulting in a total of 50 free parameters.
The following uniform prior distributions are assigned to all the knots,
\begin{equation}\label{eq:prior}
\begin{aligned}
&\ln R_{\mathrm{disc}}\mathrm{/kpc} \sim \mathcal{U}\left(\ln (0.01), \ln(8)\right), \\
&\ln \sigma_{R,0} \mathrm{/km~s^{-1}}\sim \mathcal{U}\left(\ln (10), \ln(100)\right), \\
&\ln \sigma_{z, 0} \mathrm{/km~s^{-1}}\sim \mathcal{U}\left(\ln (10), \ln(100)\right), \\
&\ln R_{\sigma, R} \mathrm{/kpc}\sim \mathcal{U}\left(\ln (1), \ln(100)\right),\\
&\ln R_{\sigma, z} \mathrm{/kpc}\sim \mathcal{U}\left(\ln (1), \ln(100)\right).
\end{aligned}
\end{equation}

To suppress oscillations in values between spline knots, we introduced a smoothing prior $P_{\mathrm{smooth}}$,
\begin{equation}\label{eq:smoothing}
    \ln P_{\mathrm{smooth}}=-\sum_{\theta}^{(R_\mathrm{disc},\sigma _{R,0},\sigma_{z,0},R_{\sigma,R},R_{\sigma,z})}\sum_{n=0}^{N_{\mathrm{knots}}-1}\frac{(\ln \theta_{n+1}-\ln \theta_{n})^2}{2(\sigma_{\theta})^2},
\end{equation}
where $\sigma_{\theta}$ is the scaling parameter, set to 0.3 for all the parameters, $\theta=(R_\mathrm{disc},\sigma _{R,0},\sigma_{z,0},R_{\sigma,R},R_{\sigma,z})$, with $\theta_{n+1}$ and $\theta_{n}$ indicating the parameter values at the $(n+1)$-th and $n$-th knots, respectively. $N_\mathrm{knots}$ is the number of knots, which was set to 10 as mentioned above. 
Adopting the smoothing prior distribution helps reduce the risk of over-fitting. 
For MCMC fitting, we used the codes developed by \citet{2024MNRAS.530.2972S}
and modified them for this application. This code uses Hamiltonian Monte Carlo with the NUTS sampler \citep{2011arXiv1111.4246H}, implemented in \texttt{NumPyro} \citep{2019arXiv191211554P}, which utilises \texttt{JAX} \citep{jax2018github} to speed up the computation.
Appendix~\ref{sec:validation} presents the verification of this model using the mock observational data, demonstrating that it recovers the ground truth within approximately 1$\sigma$ across most of the age range. This validation indicates that the uncertainties in the parameter estimates tend to increase in the age range where the number of stars is small (i.e., $\tau<2$~Gyr and $\tau>14$~Gyr). Moreover, while $\sigma_{R,0}$ and $\sigma_{z,0}$ are relatively well constrained across the full age range, $R_{\sigma,R}$ and $R_{\sigma,z}$ are more difficult to constrain especially where their values are large. This is because once a scale length exceeds the radial range spanned by the data, the radial profile becomes effectively flat within the observed region, so that the scale length cannot be tightly determined.

\begin{table*}
\centering
\caption{Summary of quasi-isothermal DF parameters described in equation \eqref{eq:DF} used in the fitting.
}\label{tab:params_sum}
\begin{tabular}{c|llccc}
\hline
 parameter $\theta$ & units & description & prior range & smoothing parameter $\sigma_{\theta}$ & figure panel \\ \hline
\Rdisc& kpc & radial scale length & (0.01, 8) &0.3&(a)\\
$\sigma_{R,0}$& km~$\mathrm{s^{-1}}$ & radial velocity dispersion at $R_\mathrm{c}=R_0$ & (10, 100) &0.3&(b)\\
$\sigma_{z,0}$& km~$\mathrm{s^{-1}}$ & vertical velocity dispersion at $R_\mathrm{c}=R_0$ & (10, 100) &0.3&(c)\\
$R_{\sigma,R}$& kpc & radial
scale length of the radial velocity dispersion profiles & (1, 100) &0.3&(d)\\
$R_{\sigma,z}$& kpc & radial
scale length of the vertical velocity dispersion profiles & (1, 100) &0.3&(e)\\
\hline
\end{tabular}

\end{table*}

\section{result}\label{sec:result}
\begin{figure*}	\includegraphics[width=\textwidth]{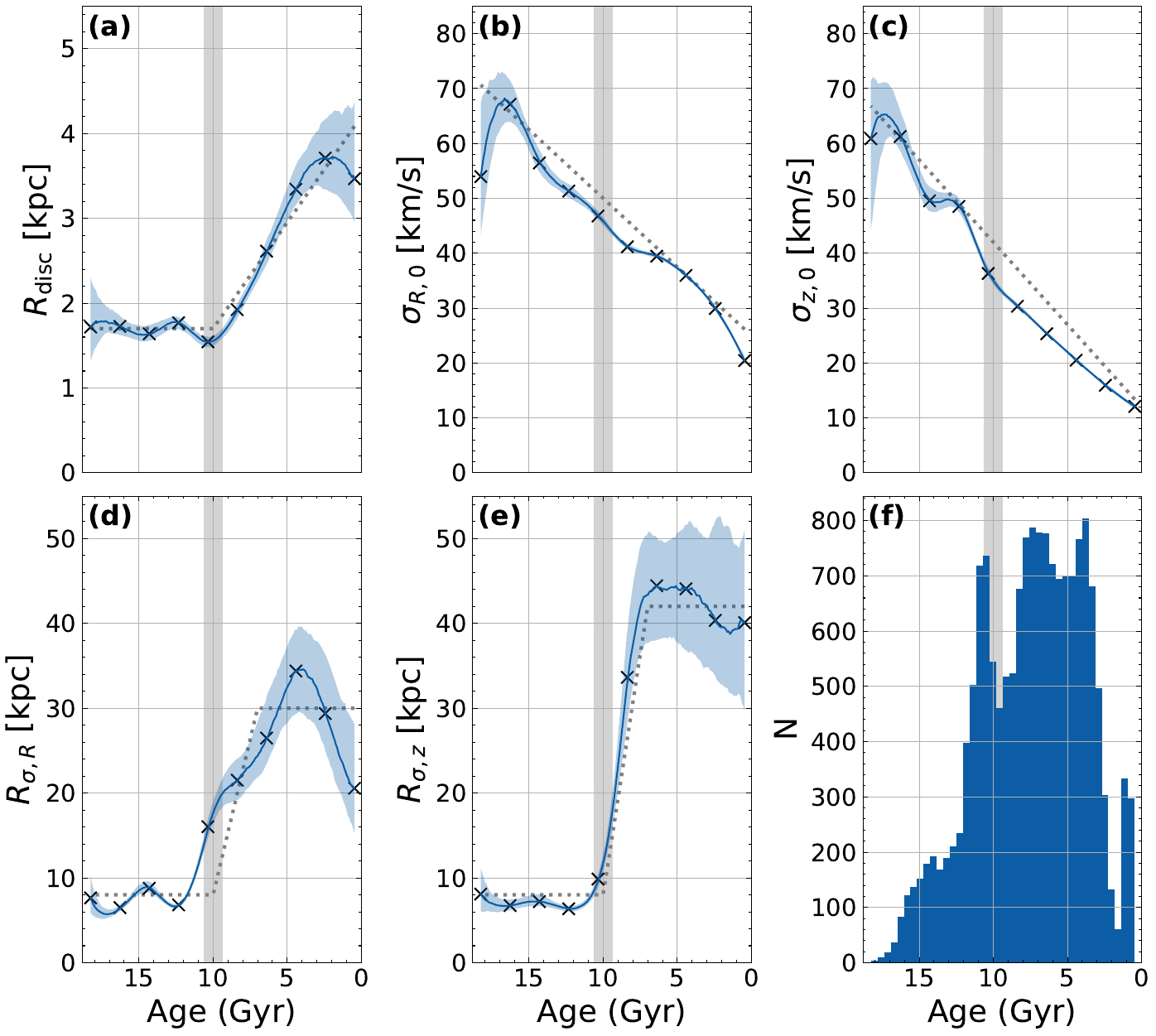}
    \caption{Posterior distributions of the splines that describe the age-dependent parameters of the quasi-isothermal DF, which are obtained after MCMC fitting. Panels (a), (b), (c), (d) and (e) correspond to the age dependence of the DF parameters $R_\mathrm{disc}, \sigma_{R,0}, \sigma_{z,0}, R_{\sigma,R}$ and $R_{\sigma,z}$, respectively, and panel~(f) shows the stellar age distribution. The blue solid lines indicate the median of the fitting results, while the shaded blue regions represent the 1$\sigma$ confidence intervals. The grey dotted lines show the parameters used in $f^{\prime}_\tau$ in equation~\eqref{eq:G}. Crosses denote the position of knots for the fitted spline function. The grey-shaded region in all the panels highlights the age around the dip in the scale length \Rdisc seen in panel~(a), which coincides with the transition from the small older, thick disc, to the growing younger, thin disc.}
    \label{fig:result:spline}
\end{figure*}

\begin{table*}
\centering
\caption{
Our best fit values (medians of the posterior distributions) for the DF parameters at the knots shown in Fig.~\ref{fig:result:spline} together with their uncertainties, which correspond to the 16th and 84th percentiles of the posterior distributions.
}\label{tab:spline_result}
\begin{tabular}{c|ccccc}
\hline
 $\tau$~[Gyr] & $R_\mathrm{disc}$~[kpc] & $\sigma_{R,0}$~[km $\mathrm{s^{-1}}$] & $\sigma_{z,0}$~[km $\mathrm{s^{-1}}$] & $R_{\sigma,R}$~[kpc] & $R_{\sigma,z}$~[kpc] \\ \hline
0.46 & $3.49^{+0.96}_{-0.56}$ & $20.46^{+0.76}_{-0.63}$ & $12.06^{+0.62}_{-0.59}$ & $20.62^{+7.29}_{-5.15}$ & $39.46^{+10.86}_{-10.25}$ \\
2.44 & $3.72^{+0.48}_{-0.37}$ & $29.94^{+0.49}_{-0.41}$ & $15.98^{+0.32}_{-0.32}$ & $29.98^{+6.80}_{-4.83}$ & $39.26^{+10.72}_{-6.32}$ \\
4.41 & $3.36^{+0.26}_{-0.19}$ & $35.94^{+0.34}_{-0.34}$ & $20.56^{+0.28}_{-0.30}$ & $34.60^{+5.18}_{-4.99}$ & $44.16^{+7.64}_{-7.30}$ \\
6.39 & $2.60^{+0.12}_{-0.15}$ & $39.51^{+0.39}_{-0.43}$ & $25.43^{+0.29}_{-0.34}$ & $26.69^{+4.45}_{-3.19}$ & $44.29^{+6.39}_{-6.36}$ \\
8.37 & $1.91^{+0.09}_{-0.09}$ & $41.20^{+0.56}_{-0.47}$ & $30.39^{+0.48}_{-0.46}$ & $21.62^{+2.45}_{-2.32}$ & $33.17^{+5.62}_{-4.76}$ \\
10.35 & $1.54^{+0.06}_{-0.06}$ & $46.92^{+0.70}_{-0.80}$ & $36.61^{+0.78}_{-0.71}$ & $15.75^{+1.33}_{-1.04}$ & $9.63^{+0.79}_{-0.79}$ \\
12.33 & $1.77^{+0.07}_{-0.08}$ & $51.40^{+1.17}_{-1.25}$ & $48.80^{+1.29}_{-1.24}$ & $6.75^{+0.27}_{-0.26}$ & $6.37^{+0.36}_{-0.35}$ \\
14.31 & $1.64^{+0.10}_{-0.09}$ & $56.78^{+2.06}_{-1.91}$ & $49.76^{+2.59}_{-2.04}$ & $8.83^{+0.67}_{-0.59}$ & $7.24^{+0.82}_{-0.57}$ \\
16.28 & $1.74^{+0.12}_{-0.11}$ & $67.17^{+4.14}_{-3.12}$ & $62.02^{+4.14}_{-3.52}$ & $6.38^{+0.44}_{-0.36}$ & $6.82^{+0.69}_{-0.61}$ \\
18.26 & $1.72^{+0.58}_{-0.46}$ & $55.81^{+11.71}_{-11.83}$ & $60.41^{+10.66}_{-13.82}$ & $7.40^{+2.18}_{-1.65}$ & $8.23^{+3.02}_{-1.99}$ \\
\hline
\end{tabular}

\end{table*}

Panels (a)-(e) of Fig.~\ref{fig:result:spline} present the fitting results for our DF model parameters as a function of stellar age, described with a cubic spline with 10 knots. Note that these fitted parameters represent present-day values for different stellar age populations in the disc, rather than the conditions at the time of their formation. The original actions at birth may not be preserved if the gravitational potential of the system has undergone rapid changes, such as a major merger. Also, radial migration \citep{2002MNRAS.336..785S} due to the bar and spiral arms leads to a change of the angular momentum of the stars.
The solid blue line shows the median of the sampled spline curves,
and the blue-shaded region shows the $\pm1\sigma$ uncertainty range.
The uncertainties in each parameter are mainly driven by the sample size.
The dashed grey line shows the age-dependent functions adopted for $f^{\prime}_\tau$, which is used for the importance sampling in equation~\eqref{eq:G}. 
We iteratively adjusted $f^{\prime}_\tau$ to be similar to our derived $f_\tau$, and our final parameters for $f^{\prime}_\tau$ are described in equation~\eqref{eq:target_params}.
Panel~(f) shows the age distribution of APOGEE stars used in our analysis.
The large number of stars younger than about 1.5~Gyr is due to the cut-off of low-mass RC stars.
Table~\ref{tab:spline_result} shows the best-fit values and uncertainties for each parameter at individual spline knots.
These are derived from the median and 16th and 84th percentiles of the posterior distribution function, respectively. Appendix~\ref{sec:resultvalidation} shows the distribution of actions for the observational data and those for the best-fit model shown with solid lines in Fig.~\ref{fig:result:spline}, divided into four age groups. Overall, the observed distributions are reproduced by the best-fit model reasonably well.

The scale length of the disc, $R_\mathrm{disc}$, remains small around 1.7~kpc for stellar populations older than 10~Gyr.
For younger populations, the disc scale length increases with decreasing stellar age.
This suggests that the age dependence of the disc scale length undergoes a transition around $\tau = 10$~Gyr.
For stars older than 10~Gyr, the small scale length is consistent with that of the high-[$\alpha$/Fe] thick disc previously reported  \citep[e.g.,][]{2012ApJ...753..148B}. The increase in \Rdisc for the younger stars demonstrates the inside-out disc growth after 10~Gyr ago. This is consistent with the growth of the low-[$\alpha$/Fe] thin disc indicated by previous studies \citep[e.g.,][]{2012ApJ...753..148B,2019ApJ...884...99F}. Our modelling robustly identifies a clear transition from the thick to the thin disc traced by stars with ages around $\tau = 10$~Gyr, marking a key epoch in the Milky Way’s disc evolution.

Both the radial and vertical velocity dispersions at the solar radius, $\sigma_{R,0}$ and $\sigma_{z,0}$, show similar trends. The velocity dispersion declines as the age decreases, consistent with previous studies \citep[e.g.,][]{2011A&A...530A.138C,2018MNRAS.475.5487S}.
We find that the vertical velocity dispersion, $\sigma_{z,0}$, declines steeply in stars with ages $12<\tau<10$~Gyr. This also indicates the transition from the thick disc, characterised by higher vertical velocity dispersion, to the thin disc, characterised by lower vertical velocity dispersion. 

Both $R_{\sigma,R}$ and $R_{\sigma,z}$ also show a significant transition traced by stars with ages around 10~Gyr.
Both parameters remain around 5-10~kpc for older stellar populations, and start to increase rapidly for stars aged around 10~Gyr.
$R_{\sigma,R}$ reaches a peak of $\sim$ 35~kpc at around 4~Gyr and $R_{\sigma,z}$ reaches a high value and stays high for $\tau\lesssim4$~Gyr.
These trends indicate that, for the thick disc population older than $\tau=10$~Gyr, radial and vertical velocity dispersions show strong radial dependence, i.e., higher velocity dispersions in the inner disc. 
In the younger thin disc population, the radial dependence of the velocity dispersions becomes weaker, and their profiles flatten. This may be linked to the flaring of young stars in the outer disc \citep[see the bottom panel of Fig.~\ref{fig:xyrz}, e.g.,][]{2014RAA....14.1406R,2017ApJ...834...27M}, and may also be consistent with a scenario where the thin disc formed from a globally cold molecular disc.
We find for the first time that $R_{\sigma,R}$ and $R_{\sigma,z}$ have an age dependence, showing trend transition with stars aged around $\tau=10$~Gyr. 

Our results show that the transition from the thick to the thin disc underlies in stars with ages around $\tau = 10 $~Gyr, as indicated by a clear change in their kinematical properties. The transitional age is indicated by vertical grey-shaded regions in Fig.~\ref{fig:result:spline}. Notably, stars at this transitional age exhibit a temporary dip in scale length reaching 1.5~kpc in panel~(a) of Fig.~\ref{fig:result:spline}. 
Appendix~\ref{sec:dipvalidation} compares the action distributions of stars with $10.5>\tau>9.5$~Gyr generated from two DF models.
The first is the best-fit DF model that includes the dip in $R_\mathrm{disc}$. The second is a comparison model without the dip, created by setting \Rdisc to 1.72~kpc, corresponding to 3$\sigma$ above the best-fit value of the knot at $\tau=10.35$~Gyr, while keeping all other parameters the same as in the best-fit DF (see Table.~\ref{tab:spline_result}). We found that the model without the dip in \Rdisc shows a slight mismatch with the observed data in the $J_R$ distribution (see details in Appendix.~\ref{sec:dipvalidation}). Hence, we consider that this dip is meaningfully inferred from the data. Interestingly, this is consistent with what is predicted by \citet{2018MNRAS.474.3629G} as introduced in Section~\ref{sec:intro}. In the next section, we discuss the implications of this dip.

\section{Discussion}\label{sec:discussion}
This section begins by examining how our model constrains the stellar density scale length, $R_\mathrm{disc}$, without density information.
We demonstrate that the dip in \Rdisc seen in Fig.~\ref{fig:result:spline} reflects a genuine signature that can be inferred from kinematic characteristics.
Next, we analyse data from an Auriga cosmological simulation, which exhibits gas disc shrinking during the transition from the thick to the thin disc formation phase \citep{2018MNRAS.474.3629G}. 
We then discuss what the \Rdisc dip reveals about the Milky Way’s formation history.

\subsection{\texorpdfstring{\Rdisc recovery from stellar kinematics}{Rdisc recovery from stellar kinematics}}\label{sec:kinematical_footprint}
\begin{figure}	\includegraphics[width=\columnwidth]{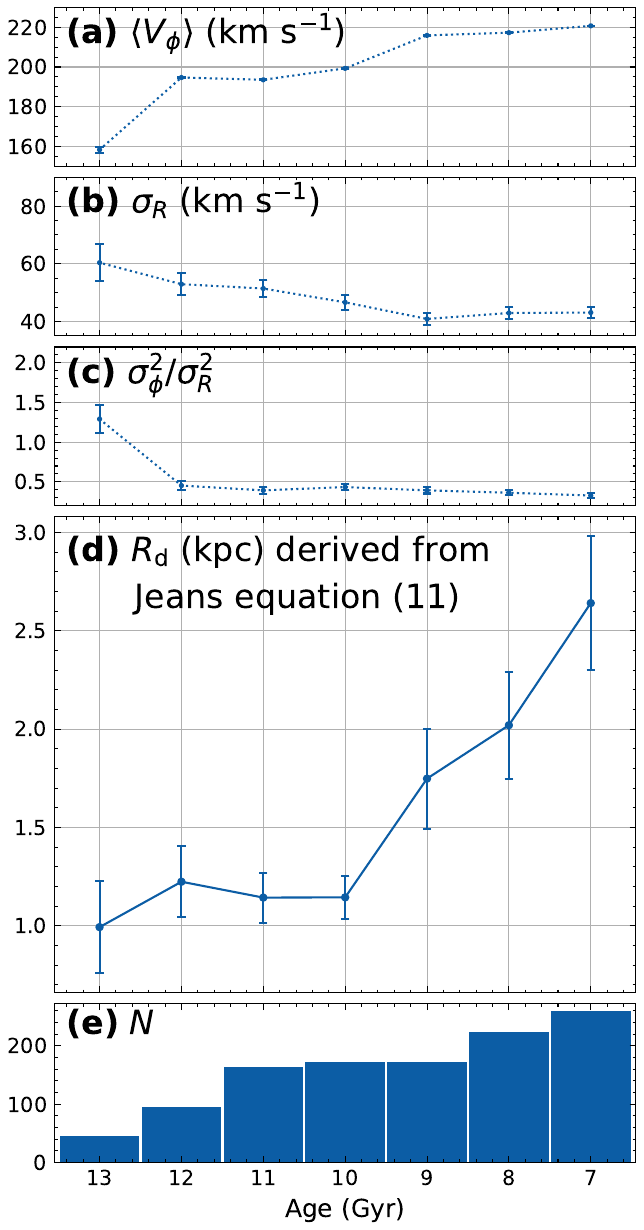}
    \caption{Measured kinematical properties (panel~a-c), the derived scale radius, $R_\mathrm{d}$, (panel~d)  and the number of stars (panel~e)  for the stars around the solar radius in different age bins.
    Solid circles with error bars connected with the dotted lines represent (a) the mean azimuthal velocity, $\langle V_\phi \rangle$, (b) the radial velocity dispersion, $\sigma_R$, and (c) the squared ratio of radial to azimuthal velocity dispersions, $\sigma_\phi^2/\sigma_R^2$, respectively. Error bars indicate the 1$\sigma$ uncertainties. Solid circles connected with the solid lines in panel~(d) show the scale length computed from these kinematic quantities via equation~\eqref{eq:jeans-eq}, with 1$\sigma$ uncertainties estimated using a Monte Carlo method shown as error bars. 
}
    \label{fig:discussion:obs_rdip}
\end{figure}

The scale length of a galactic disc is linked to stellar kinematics through the axisymmetric Jeans equation,
\begin{align}\label{eq:jeans-eq}
V_\mathrm{c}-\langle V_\phi\rangle&=\frac{\sigma^2_R}{2V_\mathrm{c}}\left[ \frac{\sigma^2_\phi}{\sigma^2_R}-1+R\left(\frac{1}{R_{\textrm{d}}}+\frac{2}{R_{\sigma,R}}\right)\right] \nonumber\\
\Leftrightarrow\quad
\frac{1}{R_\mathrm{d}}&=\frac{1}{R}\left[\frac{2 V_\mathrm{c}}{\sigma_R^2}\left(V_c-\langle V_{\phi}\rangle\right)-\frac{\sigma_{\phi}^2}{\sigma_R^2}+1\right]-\frac{2}{R_{\sigma, R}},
\end{align}
where the stellar surface density profile is described by an  exponential law as $\Sigma(R)= \Sigma_0 \exp(-R/R_\mathrm{d})$,
and velocity dispersion profiles as $\sigma_R(R)=\sigma_{R,0}\exp(-(R-R_0)/R_{\sigma,R})$. 
\Rd is the scale length of the surface density profile, and $R_{\sigma,R}$ is that of the radial velocity dispersion profile.
$V_\mathrm{c}$, $\langle V_\phi\rangle$, $\sigma_R$ and $\sigma_{\phi}$ denote circular velocity, mean rotation velocity, radial and rotational velocity dispersions at radius $R$, respectively. 
Here, $R$ refers to the current Galactocentric radius of a star. This is different from the guiding radius, $R_\mathrm{c}$, which appears in equation~\eqref{eq:DF}, i.e., \Rd in equation~\eqref{eq:jeans-eq} is defined differently from \Rdisc in equation~\eqref{eq:DF}.

To compute \Rd from equation~\eqref{eq:jeans-eq}, we focused on stars around the Galactocentric radius of the Sun, i.e. $R\sim R_0=8.275$~kpc.
We selected stars within \( |R - R_0| < 0.1 \)~kpc and \( 6.5 < \tau < 13.5 \)~Gyr, and analysed them in different age bins. Note that this binning is used only for illustrative purposes; the full DF modelling in Section \ref{sec:result} does not rely on binning and can account for age uncertainties.
The top three panels of Fig.~\ref{fig:discussion:obs_rdip} show (a)~the mean rotational velocity, \( \langle V_\phi \rangle \), (b)~the radial velocity dispersion, \( \sigma_R \)  and (c)~the ratio of the azimuthal velocity dispersion to the radial velocity dispersion, \( \sigma_\phi^2 / \sigma_R^2 \), at different age bins. Using these values and equation~\eqref{eq:jeans-eq}, we computed \Rd for each age bin, shown in panel~(d). 
Because $R_{\sigma,R}$ cannot be reliably estimated within each age bin due to the limited sample sizes and restricted radial coverage, we adopt values of $R_{\sigma,R}$ from the fitting result presented in Fig.~\ref{fig:result:spline} (Section~\ref{sec:result}).
Since we restricted our sample to a narrow range in Galactocentric radius, we fix $R = R_0$ and $V_{\mathrm{c}} = V_{\mathrm{c}}(R_0)$ in equation~\eqref{eq:jeans-eq} for simplicity ignoring the radial dependence of them.
The error bars in each panel were evaluated using the Monte Carlo method, i.e., computing these values after re-sampling the observational data using the observational uncertainties.
The number of selected stars for each bin is shown in panel~(e) of Fig.~\ref{fig:discussion:obs_rdip}.

Panel~(a) shows that \( \langle V_\phi \rangle \) increases from approximately 195~$\mathrm{km ~s^{-1}}$ at around 11~Gyr to about 220~$\mathrm{km ~s^{-1}}$ at 7~Gyr, with a transition occurring around 10~Gyr. As shown in equation~\eqref{eq:jeans-eq}, an increase of \( \langle V_\phi \rangle \) makes \Rd larger.
Similarly, in panel~(b), \( \sigma_R \) decreases from around 60~$\mathrm{km ~s^{-1}}$ at 13~Gyr to about 40~$\mathrm{km ~s^{-1}}$ after 9~Gyr. This behaviour is consistent with what we found in our cubic spline fitting in panel~(b) of Fig.~\ref{fig:result:spline}. A decrease of \( \sigma_R \) is connected to an increase of \Rd. The effect of the change in \( \sigma_R \) on \Rd is stronger the closer the value of \( \langle V_\phi \rangle \) is to $V_\mathrm{c}$, as seen from equation~\eqref{eq:jeans-eq}.
In panel~(c), a slight increase is seen in \( \sigma_\phi^2 / \sigma_R^2 \) at 10~Gyr. This leads to a larger \Rd following equation~\eqref{eq:jeans-eq}, but the contribution of this slight increase is smaller than the terms mentioned above.
Panel~(d) shows the resultant \Rd computed with equation~\eqref{eq:jeans-eq} at different age bins. 
\Rd remains small for older stars and increases for younger stars, consistent with the result shown in Fig.~\ref{fig:result:spline}.
Panel~(d) also shows a slight decrease in \Rd at $\tau=10$~Gyr, where the dip of \Rdisc is observed in panel~(a) of Fig.~\ref{fig:result:spline}. Note that, as mentioned above, \Rd and \Rdisc are defined differently, and hence it is not surprising to see that \Rd is not quantitatively the same as $R_\mathrm{disc}$. Fig.~\ref{fig:discussion:obs_rdip} shows that there is no single parameter driving the dip of $R_\mathrm{d}$. 
This suggests that the dip in \Rdisc at $\tau\sim10$~Gyr in Fig.~\ref{fig:result:spline} is obtained due to the combined trends of the mean rotation velocity and velocity dispersion properties.

The depth of the dip shown in panel~(d) in Fig.~\ref{fig:discussion:obs_rdip} is less than 0.2~kpc, which is comparable to 1$\sigma$ uncertainty, and the dip is not very clear in this figure. However, this is simply because we took a narrow radial range of the data to compute equation~\eqref{eq:jeans-eq}, which makes the error larger and the result less reliable. Note that the main aim of this section is to demonstrate that kinematical information alone is sufficient to constrain the scale length of the density distribution, not to assess the depth of the dip feature.

\subsection{Shrinking disc in Auriga cosmological simulation}\label{sec:auriga}

\begin{figure*}	\includegraphics[width=1.8\columnwidth]{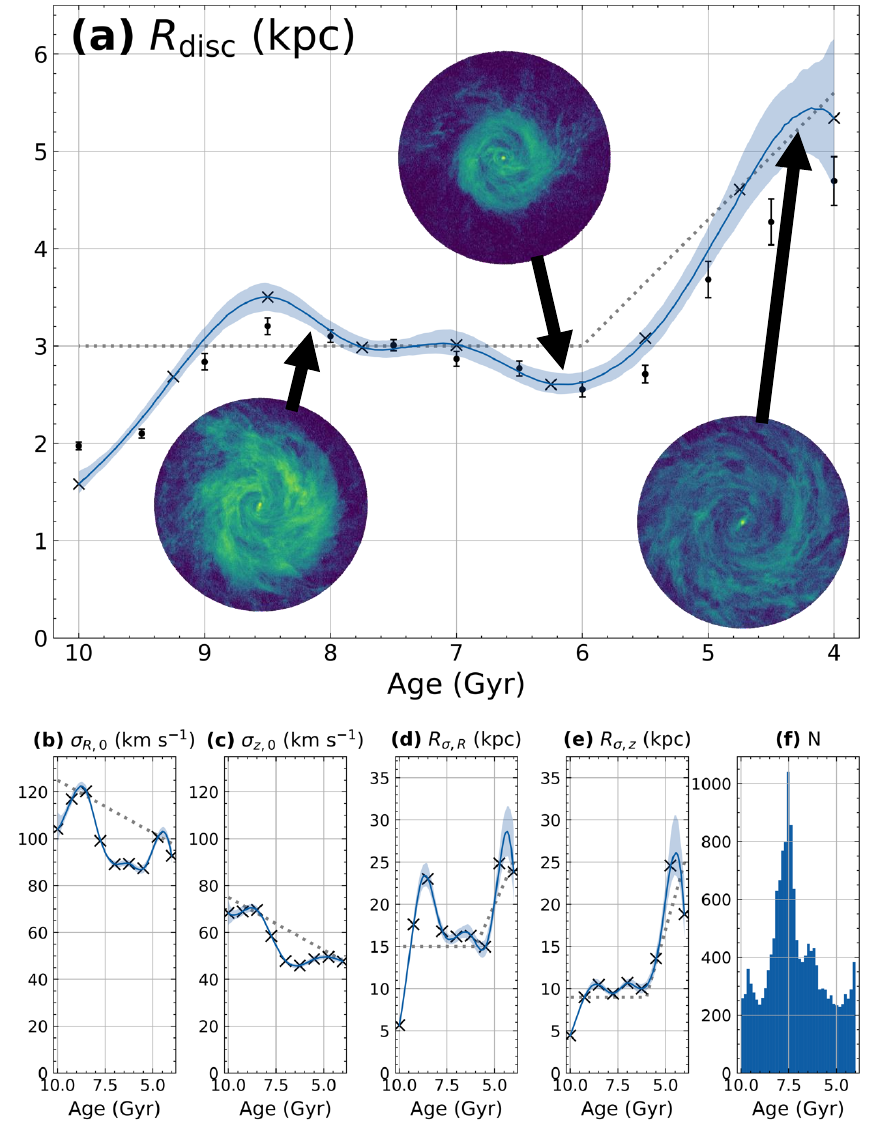}
    \caption{(Galactic disc parameters as a function of stellar age as obtained from our MCMC DF fitting for the Au~23 simulation data. Panels (a), (b), (c), (d) and (e) present the model fitting results for parameters of $R_\mathrm{disc}, \sigma_{R,0}, \sigma_{z,0}, R_{\sigma,R}$ and $R_{\sigma,z}$, respectively and panel~(f) shows stellar age distribution of the selected star particles for this fitting. The blue solid lines indicate the median of the posterior distribution function for each parameter, while the shaded blue regions represent the 1$\sigma$ confidence intervals. The thick orange solid lines show $f^{\prime}_\tau$ in equation~\eqref{eq:G}. Crosses denote the position of nine knots for the cubic spline fitting. Black dots with error bar in panel~(a) show $R_\mathrm{d}$ of the simulated galaxy in each stellar age bin measured from its surface density profile. The three snapshots shown in panel~(a) represent face-on maps of the gas disc at 4.25, 6.18, and 8.10~Gyr ago, respectively. Each image is cropped to a circular region within a radius of 20~kpc. Note that the snapshots are taken from these three different past epochs, while the parameter values shown in all panels are obtained from fitting the present-day ($z=0$) simulation data.}
    \label{fig:discussion:auriga_spline}
\end{figure*}
    
In this section, we apply our kinematic DF fitting method to a simulated galaxy from the Auriga cosmological simulation suite. The Auriga project \citep{2017MNRAS.467..179G} is a set of cosmological magneto-hydrodynamical zoom simulations of the formation and evolution of isolated Milky Way-mass galaxies,
    (mass range from $10^{12}$ to $2 \times 10^{12} \mathrm{M_\odot}$). 
    The Auriga simulations use cosmological parameters of $\Omega_m$ = 0.307, $\Omega_b$ = 0.048, $\Omega_\Lambda$ = 0.693 and Hubble constant of $H_0$ = 67.77 $\mathrm{km~s^{-1}~Mpc^{-1}}$ \citep{2014A&A...571A..16P}. 

We selected the Au23 simulation for our analysis because it closely resembles the Milky Way in terms of stellar mass and disc structure. Its stellar mass is $9.02 \times 10^{10}M_\odot$ and the radial scale length of its disc is 4.99~kpc. In addition, it is known that it exhibits a temporary shrinking of the star-forming gas disc around 6 Gyr ago, which coincides with a transition from thick to thin disc formation \citep{2018MNRAS.474.3629G}. Therefore, it is a useful test case to see whether traces of past gas disc shrinking remain in such a present-day ($z=0$) simulated galaxy as an age-dependent scale length.

    For our DF fitting we selected star particles with ages between 4 and 10~Gyr. We limited the sample to stars with $|z|<0.5$~kpc.
    Then, we randomly chose 16,000 stars within 5~kpc from the assumed observer position at $(x, y) = (-R_0, 0)$ in the simulation, where $R_0 = 8.275$~kpc corresponds to the Sun's Galactocentric radius in the Milky Way.
    The gravitational potential of the galaxy was computed with \texttt{AGAMA}'s multipole potential model assuming axisymmetry, based on the mass distributions of stars, gas and dark matter. From this potential, the circular velocity at the observer's radius was determined to be $V_\mathrm{c}(R_0) = 244.09~\mathrm{km~s^{-1}}$.

    Panels (a)-(e) of Fig.~\ref{fig:discussion:auriga_spline} show fitting results for the simulation data using cubic spline functions with nine knots indicated by crosses. As noted in Section~\ref{sec:result}, the parameters shown here do not reflect the values at the time of each star's birth (i.e., lookback time), but rather represent the present-day properties of different stellar age populations in the disc, while the three snapshots in panel~(a) illustrate the spatial gas distribution at the respective past epoch, $t_\mathrm{lookback}=4.25,6.18$ and 8.20~Gyr ago.
    The solid blue lines represent the median of the sampled spline curves from the posterior distribution function of each parameter, while the blue shaded regions indicate the 1$\sigma$ uncertainty range. The dashed grey lines show the age-dependent function for the adopted parameters of $f^{\prime}_{\tau}$. The shape of this function was determined by iteratively adjusting $f^{\prime}_{\tau}$ to be similar to the best-fitting $f_{\tau}$. 
    Panel~(f) shows the age distribution of the selected stars.

    Fig.~\ref{fig:discussion:auriga_spline} shows generally similar results to those obtained from the observational data of the Milky Way disc, and shows a clear transition of the disc structures at $\tau \sim 6$~Gyr.
    According to \citet{2018MNRAS.474.3629G}, this coincides with the epoch of the transition 
 from the high-[$\alpha$/Fe] thick disc formation phase to the low-[$\alpha$/Fe] thin disc formation phase in Au~23. Panel~(a) shows that \Rdisc for stars older than the transitional age is relatively small, around 3~kpc, and the younger disc ($\tau<6$~Gyr) shows larger $R_\mathrm{disc}$. The radial and vertical velocity dispersions (panels b and c) show the overall trend of decreasing velocity dispersion with decreasing age, though there is a large fluctuation at some ages.
    Panels (d) and (e) show that $R_{\sigma,R}$ and $R_{\sigma,z}$ also undergo a transition with stars aged around 6~Gyr, showing a smaller scale length for older stars, and an increasing scale length for younger stars. These are also similar trends to those seen in Fig.~\ref{fig:result:spline}. 

    The black dots with error bars in panel~(a) of Fig.~\ref{fig:discussion:auriga_spline} show the scale lengths measured by fitting a single exponential profile to the radial surface density of star particles in each age bin using a least-squares method. In this calculation, stars within $5<R<11$~kpc were used without applying any vertical selection.
    It should be noted that the value of \Rdisc used in the DF (eq.~\ref{eq:DF}) is not strictly identical to the scale length measured from the surface density profiles particularly when the discs are relatively warm. Although \Rdisc is formally defined as a scale length parameter in the DF, the radial density profile of stars as a function of $R$ or $R_c$ slightly deviates from an exponential profile \citep[e.g.,][]{2010MNRAS.401.2318B}. This is because the DF is described with actions, $\bm{J}$, which implicitly encode the radial dependence of the gravitational potential. This difference in the definition likely explains the difference between the scale length \Rdisc obtained from our model fitting and that derived from the surface density profiles. For example, at $\tau=7$~Gyr, our fitting result gives $R_\mathrm{disc}=3.00^{+0.14}_{-0.13}$~kpc, while the value derived from the surface density profile is $R_\mathrm{d}=2.87\pm0.08$~kpc. At $\tau=5.5$~Gyr, the corresponding values are $R_\mathrm{disc}=3.08^{+0.15}_{-0.17}$~kpc and $R_\mathrm{d}=2.71\pm0.09$~kpc, respectively.
    Despite this small difference, the overall trend of \Rdisc remains consistent with the scale length measured from the surface density profiles.
    This demonstrates that our quasi-isothermal DF fitting can successfully capture the age dependence of $R_\mathrm{disc}$, even when applied to a cosmological simulation where the galaxy does not strictly follow a quasi-isothermal DF.
    
     Notably, panel~(a) of Fig.\ref{fig:discussion:auriga_spline} shows a dip in $R_\mathrm{disc}$ around $\tau \sim 6$~Gyr, which coincides with the transition from thick to thin disc formation in Au~23.
     This dip is also seen in the scale length directly measured from the radial surface density profile at the different age bins (black dots with the error bars in panel~(a) of Fig.~\ref{fig:discussion:auriga_spline}). As reported in \citet{2018MNRAS.474.3629G}, Au~23 has a shrinking of the star-forming gas disc at $t_\mathrm{lookback}\sim6$~Gyr ago. Face-on views of the gas disc at three different look-back times (8.10, 6.18, and 4.25~Gyr from left to right) are presented in panel~(a) of Fig.\ref{fig:discussion:auriga_spline}, showing the inner 20~kpc of the galaxy. It is evident that the gas disc temporarily shrinks at 6.18~Gyr ago.
     Our results demonstrate that stars formed during the gas disc's shrinking at $t_\mathrm{lookback}\sim6$~Gyr retain a small radial scale length even at the present epoch, i.e., $z=0$. 
     Hence, drawing on these similarities between the fitting results of the Auriga simulation and observational data, the dip in \Rdisc seen in Fig.~\ref{fig:result:spline} could indicate that the Milky Way's star-forming gas disc shrank around 10~Gyr ago.

\subsection{Star-forming gas disc shrinking at the thick-to-thin disc transition}\label{sec:indication}
Our analysis suggests that the Milky Way's star-forming gas disc shrank around 10~Gyr ago, coinciding with a structural and dynamical phase transition from the thick to thin disc.
In Fig.~\ref{fig:result:spline}, stellar populations older than 10~Gyr show disc scale lengths of $R_{\mathrm{disc}}\sim 1.7$~kpc. This is consistent with previous measurements of thick disc scale length, most of which are based on star counts, e.g. 2.0~kpc by \citet{2011ApJ...735L..46B}, $1.8^{+2.1}_{-0.5}$~kpc by \citet{2012ApJ...752...51C}, $2.2\pm0.2$~kpc by \citet{2016ApJ...823...30B}, $1.9\pm0.1$~kpc by \citet{2017MNRAS.471.3057M} and 1.9~kpc by \citet{2021ApJ...912..106Y}. Importantly, our constraint is derived purely from stellar kinematics, making it complementary to star-count studies. The close agreement indicates that non-equilibrium effects are unlikely to dominate and that the gravitational potential adopted in this paper is a reasonable description of the Galaxy.

For populations younger than 10~Gyr, \Rdisc increases steadily as age decreases. This trend suggests that the thin disc grew in size smoothly. Such a pattern is consistent with the inside-out formation scenario \citep[e.g.,][]{2001ApJ...554.1044C}, which is proposed based on observed chemical abundance gradients. It is also supported by mono-abundance population analyses \citep[e.g.,][]{2012ApJ...753..148B}. Furthermore, \citet{2019ApJ...884...99F} quantified this growth, showing that the half-mass radius of the Galactic disc has expanded by approximately 43\% over the past 7~Gyr. The inside-out formation of the thin disc is consistent with observations showing that progenitors of Milky Way-sized galaxies had smaller disc sizes at higher redshifts \citep[e.g.,][]{2013ApJ...771L..35V,2024arXiv241207829T,2025MNRAS.540.3493T}. 

\citet{2018MNRAS.474.3629G} analysed the Auriga simulations and demonstrated that the shrinking of the star-forming gas disc is related to the thick-to-thin disc formation transition and the chemical bimodality in the solar neighbourhood and the outer disc of the Milky Way. They proposed that the star-forming disc shrinkage is attributed to a temporary decline in gas inflow, leading to an insufficient supply to replenish gas consumed by star formation.
In other words, when gas consumption exceeds inflow, the star-forming gas disc shrinks.

\begin{figure}	\includegraphics[width=\columnwidth]{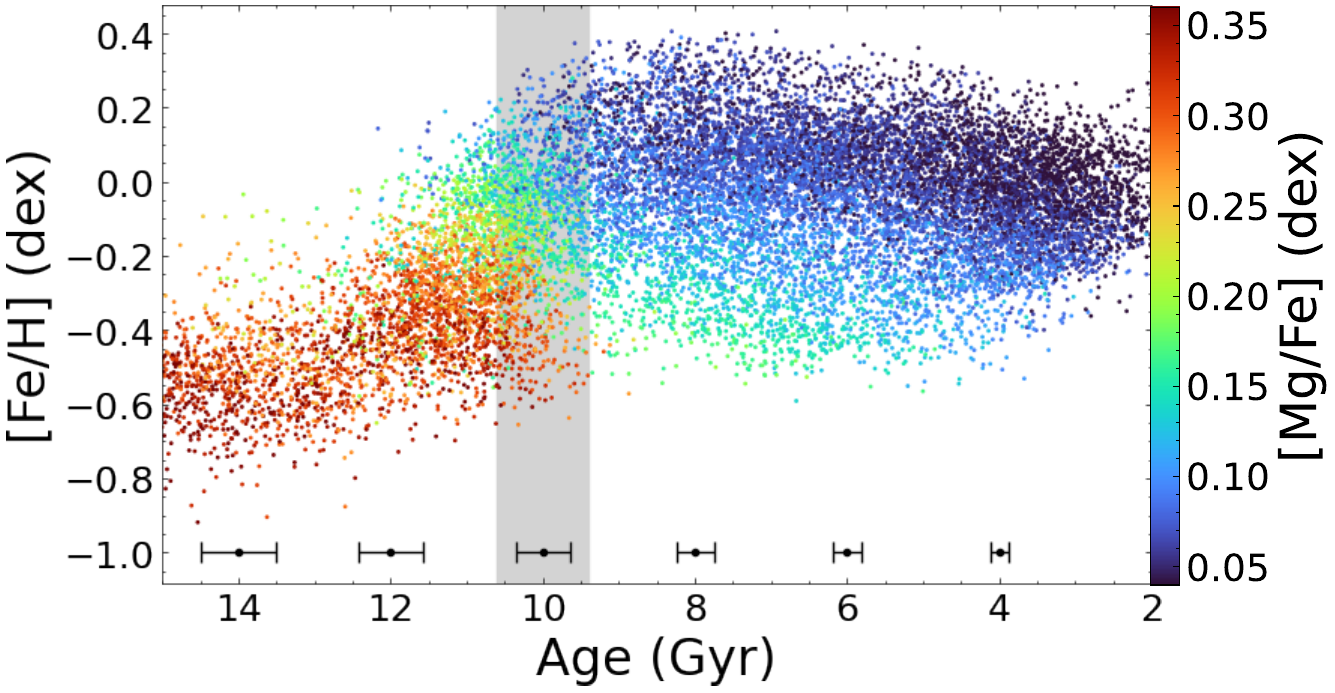}
    \caption{The age-metallicity relation coloured by [Mg/Fe]. The region shaded in grey shows the same period shown in Fig.~\ref{fig:result:spline}. Black dots with error bars represent typical 1$\sigma$ uncertainties of stars in each age. 
}
    \label{fig:discussion:age_metallicity}
\end{figure}

Fig.~\ref{fig:discussion:age_metallicity} shows the age-[Fe/H] relation in the APOGEE data used in our study, colour coded with [Mg/Fe]. 
Stars older than 12~Gyr are predominantly metal-poor ([Fe/H]$<-0.4$) with high [Mg/Fe], representing the thick disc population. In contrast, stars younger than 10~Gyr tend to have higher [Fe/H] with lower [Mg/Fe], corresponding to the thin disc. \citet{2024MNRAS.528L.122C} referred to the period $12>\tau>10$~Gyr as the GGS phase, where the metallicity rapidly increases with decreasing age, and [Mg/Fe] decreases at the same period. By comparing APOGEE data with Auriga simulations, \citet{2024MNRAS.528L.122C} suggested that this chemical transition was driven by the gas-rich GSE merger.
The gas inflow from the GSE merger likely induced a starburst in the Galactic disc, initially boosting [Mg/Fe], followed by a rise in [Fe/H] and a fall in [Mg/Fe] \citep[e.g.,][]{2007ApJ...658...60B}. 
The grey-shaded region in Fig.~\ref{fig:discussion:age_metallicity} marks the same epoch as in Fig.~\ref{fig:result:spline}, highlighting the phase transition epoch with the temporary disc shrinking in the Milky Way. This period coincides with the final stage of the GGS phase. One possible scenario is that the intense starbursts caused by the GSE could have exceeded the gas supply and lead to the shrinking of the star-forming gas disc.

Another interesting explanation is the transition from cold-mode to hot-mode gas accretion. 
Theoretical work by \citet{2018Natur.559..585N} proposed that high-$[\alpha/\mathrm{Fe}]$ stars formed rapidly through cold-mode gas accretion, followed by the formation of low-$[\alpha/\mathrm{Fe}]$ stars under hot-mode accretion. In the cold mode, which dominates in lower-mass galaxies, gas flows efficiently along cosmic filaments into the cold star-forming gas disc without experiencing shock heating, since their virial temperatures are low. In contrast, hot-mode accretion dominates in more massive galaxies. In the hot mode dominant galaxies, inflowing gas is first heated to the temperature of the hot halo. The present-day Milky Way is thought to be massive enough for hot-mode accretion, which may be essential for the thin disc formation \citep[e.g.,][]{2005MNRAS.363....2K,2018MNRAS.474.3629G,2022MNRAS.514.5056H}. Thus, the Milky Way would have undergone this transition at some point in the past, eventually reaching the hot mode today. \citet{2018MNRAS.474.3629G} showed that the shrinking of the star-forming gas disc at the onset of thin disc formation occurred when the Milky Way became massive enough for hot-mode gas accretion to dominate, reducing the gas supply to the central cold gas disc.

Building on the insights from the simulation study in \citet{2020MNRAS.497.1603G} and the timing coincidence between the end of the GGS phase and the scale length dip, our results may suggest the following potential scenario to explain the shrinking of the gas-disc at the transition from the thick to thin disc formation phase in the Milky Way: (1) before the GSE merger, the Milky Way was still relatively low in mass, allowing the formation of the thick disc through intense star formation and occasional gas-rich mergers under cold-mode gas accretion, which maintained the formation of high-[$\alpha$/Fe] thick disc stars \citep{2004ApJ...612..894B,2005ApJ...630..298B,2012MNRAS.426..690B}. (2) The gas-rich nature of the GSE merger \citep{2020MNRAS.497.1603G,2024MNRAS.528L.122C} likely triggered an intense starburst, leading to rapid consumption of the available gas.
At the same time, the GSE merger may have been substantial enough to increase the total mass of the Milky Way and triggered a transition from cold-mode to hot-mode gas accretion. Alternatively, this transition may have resulted from a combination of factors, with the GSE being only one of them. The total mass of GSE is subject to considerable debate, with estimates ranging from $\sim 10^{8}$ to $10^{11} M_\mathrm{\odot}$, corresponding to $\sim10-50\%$ of the Milky Way’s progenitor mass at the time, depending on the study \citep[e.g., ][]{2018Natur.563...85H,2021ApJ...923...92N,2023MNRAS.523.1565B,2023MNRAS.526.1209L}. The critical halo mass threshold between hot- and cold-modes is estimated approximately $10^{11.4}$ to $10^{12} M_\mathrm{\odot}$ \citep{2005MNRAS.363....2K,2006MNRAS.368....2D}.
(3) The transition to hot-mode accretion reduced the inflow of cold gas, just after the rapid consumption of the gas due to the GSE merger, limiting the Galaxy's replenishment of its star-forming gas and thereby accelerating the depletion of the gas reservoir. This process may have led to the shrinking of the cold gas disc, leaving an imprint that is now observed as the dip in \Rdisc traced by stars aged around $\tau\sim10$~Gyr as in Fig.~\ref{fig:result:spline}. (4) The subsequent formation of the thin disc proceeded in an inside-out manner, as indicated by the steady increase of \Rdisc for younger stellar populations shown in Fig.~\ref{fig:result:spline}.

A star formation quenching in the solar-neighbourhood of the Galactic disc has been proposed in the literature \citep[e.g.,][]{2000A&A...358..671G}. This can explain the observed bimodality in the [$\alpha$/Fe]–[Fe/H] distribution of solar-neighbourhood stars \citep[e.g.,][]{2016A&A...589A..66H,2014ApJ...781L..31S,2015A&A...578A..87S}. To account for this feature, chemical evolution models such as the two-infall model have been proposed, in which a temporary suppression in star formation is caused by a delay between two episodes of gas accretion \citep[e.g.,][]{1997ApJ...477..765C,2015IAUGA..2255326C}.
The halt of star formation is thought to have occurred during the transitional epoch between the formation of the thick and thin discs \citep[e.g.,][]{2024A&A...690A.208S}, which corresponds to around 10~Gyr ago on the stellar age scale used in this study. 
In our study, we identify a dip in the disc scale length $R_\mathrm{disc}$ around $\tau \sim 10$~Gyr, which aligns with this thick-to-thin disc transition period. If the gas disc during the thick disc formation phase was initially extended beyond the solar radius and shrank to within the solar radius ($R_0$) at the transitional epoch, star formation would have been temporarily halted in the outer disc—including at $R_0$, while continuing in the inner disc, as suggested in \citet{2018MNRAS.474.3629G}. This "outer-disc quenching" scenario may provide an explanation for the observed chemical bimodality: star formation near the solar radius was paused when the gas disc shrank, and resumed once fresh gas was steadily accreted and the disc regrew in an inside-out manner, eventually reaching $R_0$. In this scenario, star formation at the solar radius naturally ceased due to the temporarily reduced size of the star-forming disc. 

While the scenario above offers a more physically motivated explanation for the pause in star formation inferred from solar-neighbourhood stars, it is not the only possible interpretation. To test this scenario, further modelling will be required, including the effects of radial migration and the kinematical heating due to the merger, which may have played a key role in reshaping the stellar disc’s scale length. 
In fact, the dip of scale length we found in Fig.~\ref{fig:result:spline} may not look drastic to infer such outer-disc quenching. However, our analysis is the present-day scale length of the stars formed during this transition period, whose angular momentum distribution must be significantly modified by radial migration and radial mixing due to the heating. 
For example, thick disc stars may have formed initially at smaller radii and then moved outward as the thin disc grew through inside-out formation, receiving angular momentum from the thin disc or being dynamically heated by minor mergers and/or the bar \citep[e.g.,][]{2012MNRAS.426..690B}.
The upper panel of Fig.~\ref{fig:discussion:auriga_spline} demonstrates that the gas disc experienced a significant shrinking of about one-third of its size between $t_\mathrm{lookback} = 8.10$ and 6.18~Gyr ago. In contrast, the difference in the stellar scale length at $z=0$ is only about one tenth. This suggests that the small dip observed in the present-day scale length may reflect a larger difference in the star-forming gas disc at the birth epoch of those stars.

\citet{2025ApJ...983L..10Z} compared the age-metallicity relation from \Gaia data with simulations including a slowing bar, and proposed that the age-metallicity relation shows two sequences: a higher-[Fe/H] sequence formed by stars that migrated radially due to the slowing bar, and a lower-[Fe/H] sequence reflecting the chemical evolution of stars formed locally.
The sequence of local chemical evolution for stars with guiding radii around $R_g \sim 8$ kpc appears to begin approximately 8 Gyr ago, at a metallicity of [Fe/H] $\sim -0.75$. Note that their age scale is different from our age scale. This trend is consistent with our proposed scenario of inside-out thin disc formation, after the star forming gas disc shrinking. Star formation at $R \sim 8$~kpc likely began when the gas disc grew outward and reached that radius, fuelled by the accretion of low-[Fe/H] gas from the hot halo gas. Interestingly, in the scenario proposed by \citet{2025ApJ...983L..10Z}, stars formed in the inner disc can migrate outward to $R\sim$8~kpc due to the growth of the bar. This implies that the star-forming disc during the thick disc formation phase did not need to extend as far as $R\sim$8~kpc, and that the star formation at this radius began only when the thin disc had grown large enough to reach it.

Clumpy early disc formation has also been suggested to explain the early thick disc formation \citep{1998Natur.392..253N,2009ApJ...707L...1B,2016MNRAS.456.2052I}. In high-redshift galaxies, gas-rich discs become gravitationally unstable, leading to the formation of massive, high-density gas clumps. Within these clumps, star formation proceeds rapidly and can drive the production of high-[$\alpha$/Fe]  populations. Previous studies comparing Milky Way observations with simulations have suggested that clump formation can reproduce key features of the disc such as the [$\alpha$/Fe] bimodality \citep[e.g.,][]{2019MNRAS.484.3476C,2020MNRAS.492.4716B}. 
Clumpy disc scenario generally predicts that the high- and low-[$\alpha$/Fe] discs evolve in parallel rather than sequentially \citep[e.g.,][]{2021MNRAS.502..260B}, which could be a distinctive feature from the scenario of the thick to thin disc transition due to the GSE merger highlighted in this paper. However, our model focuses on the age dependence of the kinematical structure of the Galactic disc, and cannot distinguish these scenarios. Fitting chrono-chemodynamical data with the model including the parametric descriptions of the impacts of GSE-like merger, clumpy disc formation, the bar and spiral arms would be interesting to further disentangle  the formation history of the Milky Way, though it would be challenging to construct such model.

It is important to keep in mind the uncertainties in stellar age estimates when considering any scenario or assessing causal links between different events. Fig.~\ref{fig:discussion:age_metallicity} shows typical age uncertainties for the stars used in this study, indicated by error bars at several stellar ages. 
Around $\tau = 10$~Gyr, where the transition from the thick to thin disc occurs and a dip in the scale length is observed, the typical age uncertainty is about 0.35~Gyr. 
Additionally, the spline fitting in Fig.~\ref{fig:result:spline} uses knot intervals of approximately 2~Gyr (see Table~\ref{tab:spline_result}). Therefore, the effective age resolution of our analysis is around 2~Gyr, and trends on finer age scales (i.e., $<\sim$2~Gyr) cannot be reliably discussed.
Moreover, the stellar ages used in this study are based on the BINGO age trained with asteroseismic age \citep{2021A&A...645A..85M}. Its absolute scale differs from other methods, and the relation between different age estimates is not necessarily linear. A key challenge for future work is to calibrate the age scales across different estimation techniques.
    
\section{Summary}\label{sec:summary}
In this study, we analysed the stellar age-kinematics relation of 16,617 red giant stars from APOGEE DR17, using age estimates from \citet{2024MNRAS.528L.122C}, cross-matched with \Gaia DR3. 
Following the approaches of \citet{2023MNRAS.521.1462Z} and \citet{2024MNRAS.530.2972S}, we modelled the stellar kinematic distribution using a dynamical action-based distribution function parametrized by cubic spline curves to describe the age dependence of the disc parameters. 

We provide the first DF model that smoothly tracks structural and kinematic properties across the entire stellar age range in the Milky Way disc. Our results reveal a clear kinematic transition at age $\sim 10$~Gyr, marking the shift from the thick disc to the thin disc population. This transition is accompanied by a decrease in the disc scale length, followed by inside-out growth of the thin disc. 
This is the first observational indication of such a temporary shrinking at the transition period between high-[$\alpha$/Fe] thick disc and low-[$\alpha$/Fe] thin disc. Importantly, we also find for the first time that $R_{\sigma,R}$ and $R_{\sigma,z}$ exhibit a clear age dependence, with a transition in their trends occurring at age $\sim 10$~Gyr. How the transition of these parameters is connected to the formation history of the Galactic disc remains to be clarified. These trends may reflect processes such as flaring and dynamical heating in the outer disc. This should be further tested with numerical simulations in future work.

We applied the same fitting approach to an Auriga cosmological simulation, Au~23, which shows the temporary shrinking of the star-forming cold gas disc at the period of the transition from the high-[$\alpha$/Fe] thick to the low-[$\alpha$/Fe] thin disc formation \citep{2018MNRAS.474.3629G}. 
Our method recovers the short radial scale length of the stars formed during the cold gas-disc shrinking period. This demonstrates that past gas disc shrinking can leave imprints on present-day stellar kinematics, which can be recovered through our DF fitting. 

Hence, we suggest that the small disc found for stars aged around $\tau=10$~Gyr reflects a temporary shrinking of the Milky Way's star-forming gas disc at that time. This event coincided with the end of the GGS phase, which was likely triggered by the gas-rich GSE merger \citep{2024MNRAS.528L.122C}. Notably, this timing also corresponds to the onset of the low-[$\alpha$/Fe] thin disc formation (Fig.~\ref{fig:discussion:age_metallicity}).
Drawing on our results and previous simulation studies \citep[e.g.,][]{2018MNRAS.474.3629G,2020MNRAS.497.1603G}, we outline a possible scenario that may explain our findings.
Before the GSE merger, the Milky Way was low enough mass to maintain predominantly cold-mode gas accretion, driving high star formation rates and fed by frequent gas-rich mergers, as expected at high redshift in a $\Lambda$CDM Universe \citep[e.g.,][]{2004ApJ...612..894B}. The gas-rich GSE merger then triggered an intense starburst \citep[GGS phase;][]{2024MNRAS.528L.122C} and rapid gas shrinking of the disc. 
At the same time, a transition from cold- to hot-mode gas accretion may have occurred, possibly triggered by the increase in the Milky Way’s mass, due to the significant mass added by GSE merger.
The transition to hot-mode accretion slowed the supply of gas to the cold gas disc, just as the cold gas was being rapidly consumed by the starburst. This combination could have caused the temporary shrinking of the star-forming gas disc.
As a result, the thin disc formation started from a gas disc smaller than the thick disc. Subsequently, the thin disc grew inside-out, fuelled by smooth and gradual gas accretion from the hot halo. Over time, gas with progressively higher angular momentum accreted, driving the disc’s outward expansion \citep[e.g.,][]{2012MNRAS.419..771B}. 

We caution that stellar age estimates carry uncertainties that affect the interpretation of features such as the dip in scale length observed in stars with ages around $\tau\sim10$~Gyr. The BINGO inferred age used here may differ in scale from ages derived by other methods. Calibrating age scales across different techniques is essential for comparing the timing of key events and building a coherent picture of the Milky Way's formation history.

Recently, \citet{2024MNRAS.531.1520M} suggested that the GSE merger can trigger bar formation. 
Hence, comparing key epochs in the Milky Way's evolution—including the GGS period triggered by the GSE merger \citep{2024MNRAS.528L.122C}, the thick-thin disc transition epoch revealed in our study and the bar formation period suggested by \citet{2024MNRAS.530.2972S}—is essential for understanding the comprehensive picture of the Milky Way's history. However, the age scale of our study based on \citet{2021A&A...645A..85M} is different from the age scale used in \citet{2024MNRAS.530.2972S} which relies on the calibrated Mira variable age-period relation in \citet{2023MNRAS.521.1462Z}. We will carefully calibrate these age scales and investigate how bar formation influences the transition between thick and thin disc formation phases in future work.

\section*{Acknowledgements}
We thank the anonymous referee for the careful review and helpful comments.
This work is a part of MWGaiaDN, a Horizon Europe Marie Sk\l{}odowska-Curie Actions Doctoral Network funded under grant agreement no. 101072454 and also funded by UK Research and Innovation (EP/X031756/1). This work was also partly supported by the UK's Science \& Technology Facilities Council (STFC grant ST/S000216/1, ST/W001136/1). JLS acknowledges the support of the Royal Society (URF\textbackslash R1\textbackslash191555; URF\textbackslash R\textbackslash 241030).

This work has made use of data from the European Space Agency (ESA) mission \Gaia  (\url{https://www.cosmos.esa.int/gaia}), processed by the \Gaia Data Processing and Analysis Consortium (DPAC, \url{https://www.cosmos.esa.int/web/gaia/dpac/consortium}). Funding for the DPAC has been provided by national institutions, in particular the institutions participating in the \Gaia Multilateral Agreement.

This work has made use of data from the Apache Point Observatory Galactic Evolution Experiment (APOGEE), which is part of the Sloan Digital Sky Survey (SDSS). Funding for the Sloan Digital Sky Survey IV has been provided by the Alfred P. Sloan Foundation, the U.S. Department of Energy Office of Science, and the Participating Institutions. SDSS acknowledges support and resources from the Center for High-Performance Computing at the University of Utah. The SDSS web site is \url{www.sdss4.org}.

SDSS is managed by the Astrophysical Research Consortium for the Participating Institutions of the SDSS Collaboration including the Brazilian Participation Group, the Carnegie Institution for Science, Carnegie Mellon University, Center for Astrophysics | Harvard \& Smithsonian (CfA), the Chilean Participation Group, the French Participation Group, Instituto de Astrofísica de Canarias, The Johns Hopkins University, Kavli Institute for the Physics and Mathematics of the Universe (IPMU) / University of Tokyo, the Korean Participation Group, Lawrence Berkeley National Laboratory, Leibniz Institut für Astrophysik Potsdam (AIP), Max-Planck-Institut für Astronomie (MPIA Heidelberg), Max-Planck-Institut für Astrophysik (MPA Garching), Max-Planck-Institut für Extraterrestrische Physik (MPE), National Astronomical Observatories of China, New Mexico State University, New York University, University of Notre Dame, Observatório Nacional / MCTI, The Ohio State University, Pennsylvania State University, Shanghai Astronomical Observatory, United Kingdom Participation Group, Universidad Nacional Autónoma de México, University of Arizona, University of Colorado Boulder, University of Oxford, University of Portsmouth, University of Utah, University of Virginia, University of Washington, University of Wisconsin, Vanderbilt University, and Yale University.
\section*{Data Availability}
The data underlying this article will be shared on reasonable request to the corresponding author.
We have used simulations that are part of the Auriga Project public data release \citep{2024MNRAS.532.1814G} available at \url{https://wwwmpa.mpa-garching.mpg.de/auriga/data}.



\bibliographystyle{mnras}
\bibliography{example} 




\appendix

\section{Mock data validation}\label{sec:validation}
We here demonstrate that our model-fitting method described in Section~\ref{sec:modelling} can recover the known parameters of mock galaxy data. We used \texttt{AGAMA} to generate mock data with the same spatial and age distribution as our observational data, but assigning kinematics based on a quasi-isothermal DF (eq.~\ref{eq:DF}) with the assumed parameter values as a function of age.
Then, we fitted the mock data using a quasi-isothermal DF, modelling the age dependence of the parameters with cubic splines. To generate mock data we used the same galactic potential and solar position as mentioned in Section~\ref{sec:modelling}.

The age-dependent parameters for the DF used to generate mock data are set as:
\begin{equation}
\begin{aligned}
R_{\mathrm{disc}}&=
\begin{cases}-0.25 \times \tau +4.2 & \text { if } \tau \leq 10, \\
1.7& \text { otherwise },\end{cases} \\
\sigma_{R,0}&=2.5\times\tau+25,\\
\sigma_{z,0}&=3\times\tau+12,\\
R_{\sigma,R}&=\begin{cases}30 & \text { if } \tau \leq 7, \\
-\frac{22}{3}\times\tau+\frac{244}{3}& \text { if } 7 <\tau \leq 10,\\
8& \text { if } \tau>10,\end{cases} \\
R_{\sigma,z}&=\begin{cases}42 & \text { if } \tau \leq 7, \\
-\frac{34}{3}\times\tau+\frac{364}{3}& \text { if } 7 <\tau \leq 10,\\
8& \text { if } \tau>10,\end{cases} \\\label{eq:target_params}
\end{aligned}
\end{equation}
where $\tau$ is in Gyr, $R_\mathrm{disc}, R_{\sigma,R}$ and $R_{\sigma,z}$ are in kpc and $\sigma_{R,0}$ and $\sigma_{z,0}$ are in $\mathrm{km~s^{-1}}$.
Based on the stellar age of each star in the observational dataset, we determined five parameters for the DF using equation~\eqref{eq:target_params}. We generated 10,000 sample stars for each observed star using its corresponding DF defined by the parameters described above. Among these samples, we selected the one whose ($R$, $z$) position was closest to that of the corresponding observed star.
Next, we replaced the azimuthal angle of the selected data point with that of the observed star, keeping its $R$ and $z$ coordinates fixed. The stellar age of the observed star was then assigned to the selected point. This procedure allows us to incorporate observational uncertainties in a manner consistent with the real data. We note that the observed ($R, z$) positions include measurement errors, while the matched samples are drawn from error-free distributions.
By repeating this process for all stars in the observational data, we generated a mock dataset containing the same number of stars as the observational data. The mock data have a similar spatial distribution and an identical stellar age distribution to the observational data. Thanks to the sampling method above, the velocities of the generated mock data follow the DF whose parameters vary according to equation~\eqref{eq:target_params}.

The mock data were converted into observable coordinates, $(\ell, b, \varpi, \bm{\mu}, v_{\|})$.
We then assigned the observational errors for each star, including the uncertainty in stellar age, to the corresponding mock data points.
We used the log-uniform priors described in equation~\eqref{eq:prior} and the smoothing priors given in equation~\eqref{eq:smoothing}. For the importance sampling, we assumed that $f^\prime_\tau$ in equation~\eqref{eq:G} shares the same age dependence as that of the target mock data described in equation~\eqref{eq:target_params}.

\begin{figure*}	\includegraphics[width=\textwidth]{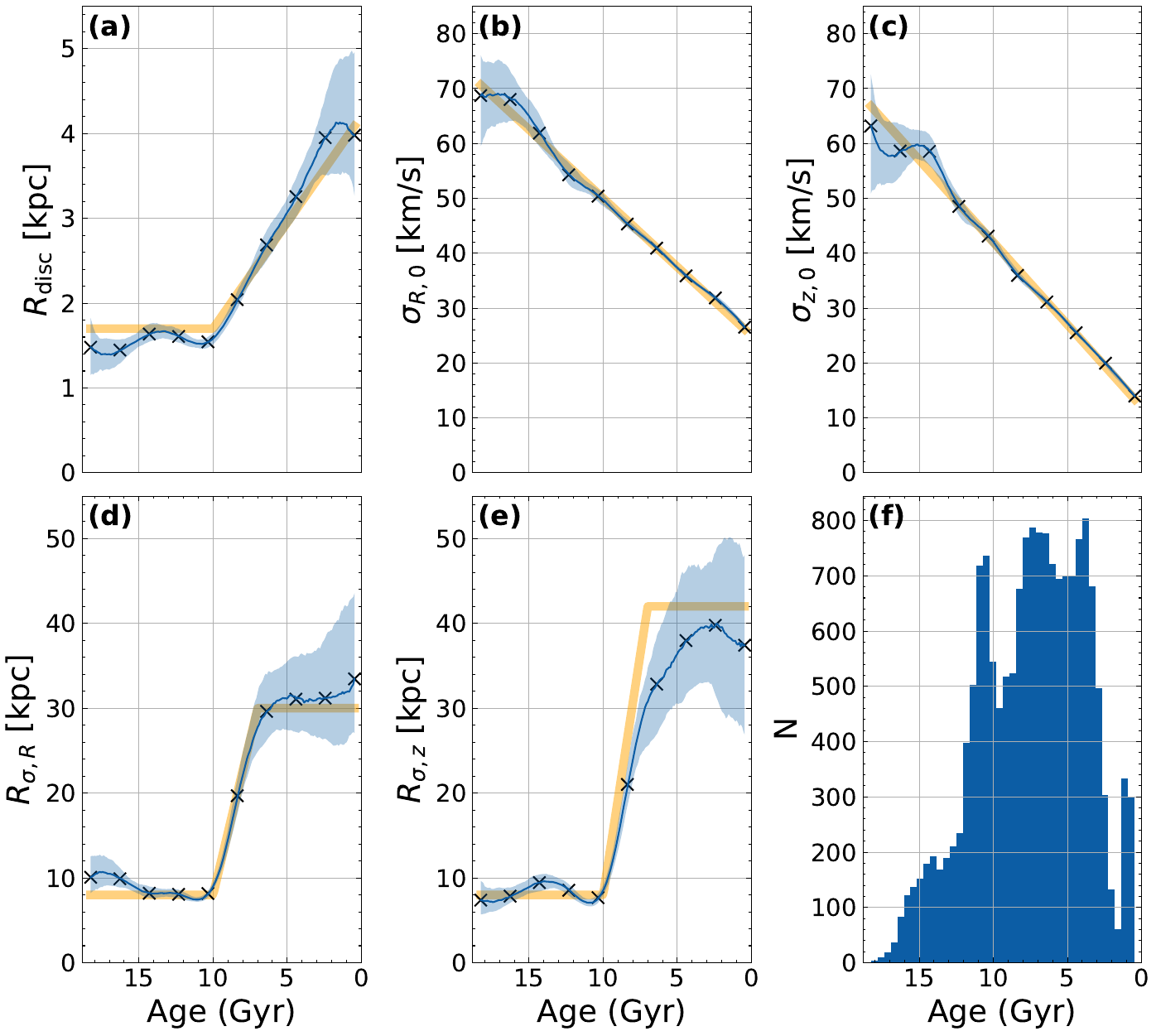}
    \caption{Results of model validation with a mock observational data. Panels (a), (b), (c), (d) and (e) present the model fitting results for parameters of $R_\mathrm{disc}, \sigma_{R,0}, \sigma_{z,0}, R_{\sigma,R}$ and $R_{\sigma,z}$ as a function of age, respectively, and panel~(f) shows stellar age distribution of the mock data. The blue solid lines indicate the median probability of the fitting results for each parameter, while the shaded blue regions represent the 1$\sigma$ confidence intervals. The thick orange solid lines show the age-dependent functional forms for each parameter, used to generate the mock data, i.e. true parameter values. Crosses denote the position of 10 evenly spaced knots set for the cubic spline fitting of the age dependence of each parameter.
}
    \label{fig:modeling:validation}
\end{figure*}

Panels (a)-(e) in Fig.~\ref{fig:modeling:validation} show the results of our fitting with 10 knots indicated by crosses.
The solid blue line shows the median of the sampled spline curves from the posterior distribution of the parameters,
while the blue-shaded region shows the 1$\sigma$ dispersion range.
The thick orange lines show the true parameters used to generate the target mock data, as described in equation~\eqref{eq:target_params}.
Panel~(f) shows the age distribution of the mock data.
The stellar ages in the mock sample are taken from the observational data so that the age distribution of the mock data is identical to that of the observational data.

Fig.~\ref{fig:modeling:validation} shows that our fitting successfully recovers the true parameter values within approximately 1$\sigma$ over most of the age range.
Panel (a) of Fig.~\ref{fig:modeling:validation} shows that \Rdisc fluctuates at $\tau>10~$Gyr even though there are no corresponding features in the mock data. At 10 Gyr, \Rdisc is smaller than at 12 Gyr and shows a dip similar to that seen in the observed data. However, the depth of this dip is less than 1$\sigma$, while the dip in the real data reaches approximately 4$\sigma$ (Fig.~\ref{fig:result:spline}).
The uncertainties in the estimates of each parameter are mainly dominated by the small number of data points. The number of stars is lower, particularly at $\tau \lesssim 2$~Gyr and $\tau \gtrsim 14$~Gyr, as shown in panel~(f). In this age range, all parameters exhibit relatively large uncertainties.
The error ranges for $\sigma_{R,0}$ and $\sigma_{z,0}$ appear small across all ages simply because these parameters are inherently easier to constrain with relatively high accuracy. 
Although the uncertainties remain small even in the age range $\tau \lesssim 2$~Gyr, where the number of stars is low, this is a result of both the smaller estimated parameter values and their tighter constraints.
The uncertainties in $R_{\sigma,R}$ and $R_{\sigma,z}$ increase sharply for stars younger than 7~Gyr, despite the large number of stars around that age range. This is likely because the larger values of these parameters reduce the radial contrasts in velocity dispersions, making the fitting less stable.

It should be noted that the results exhibit some variation due to the stochastic nature of the sampling process to generate the mock data from the limited number of observed stars.
The validation result shown in Fig.~\ref{fig:modeling:validation} represents the results from one realisation of the mock data. We have also generated several mock data using different random sampling from the above-mentioned methodology, keeping the DF same. We confirm that the results are consistent with each other roughly within 1$\sigma$.
Nevertheless, overall, the parameter recoveries achieved by our fitting are good, and the associated uncertainties are statistically meaningful.
This demonstrates that the model can robustly capture age-dependent trends in Galactic parameters.

\section{Verification of our best-fit model}\label{sec:resultvalidation}
To verify the fitting results presented in Section~\ref{sec:result}, we constructed mock samples based on the best-fitting age-dependent DF parameters (shown in Fig.~\ref{fig:result:spline}). 
For each observed star, we used \texttt{AGAMA} to get the velocity DF $f(\bm{v}|\bm{x}_i)$ at its position $\bm{x}_i$, and generated 100 velocity samples from this DF. Random shifts were then applied to the parallax, proper motions, and line-of-sight velocity of each sample, based on the corresponding observational uncertainties. Finally, we computed the actions for all generated mock stars. 

\begin{figure*}	\includegraphics[width=\textwidth]{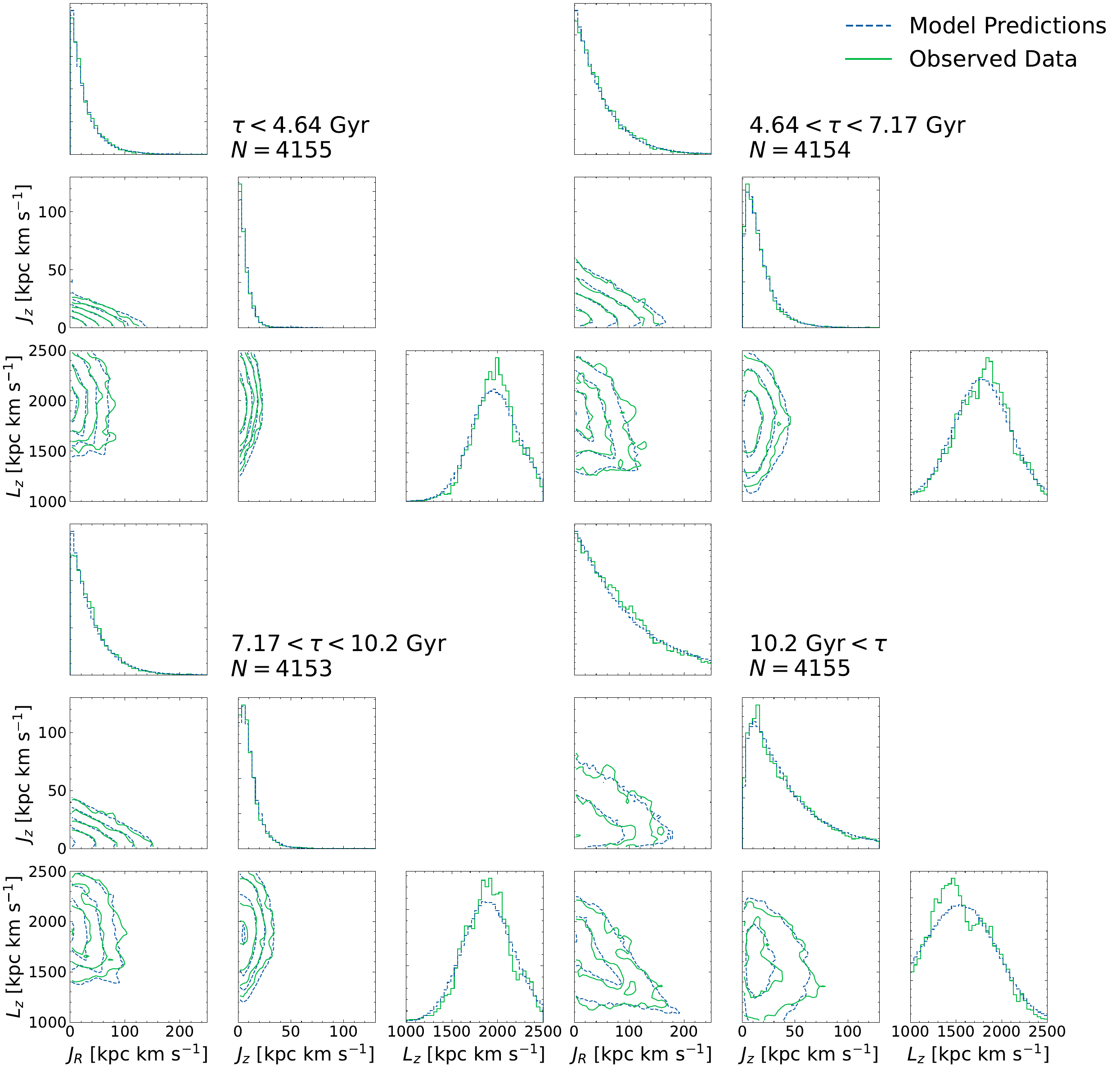}
    \caption{Comparison of the action-space distributions between our best-fit DF model (blue dashed lines), shown in Fig.~\ref{fig:result:spline}, and the observed data (green solid lines). For clarity, the sample is divided into four different stellar age groups. Overall, the model reproduces the observed distributions well.
}
    \label{fig:appendix:4panels_actions}
\end{figure*}

The comparison of the action distribution between the best-fit model and the observational data are shown in Fig.~\ref{fig:appendix:4panels_actions}. For clarity, the samples are divided into four groups by age with almost equal sample sizes. The action distributions exhibit distinct structures in different age groups, which is consistent with the result shown in Section~\ref{sec:result}.
We find generally good agreement between our best-fit DF model and the observational data across all age ranges.
The histogram of angular momentum, $L_\mathrm{z}$ appears to be less consistent than the other two physical quantities. These may be due to resonances caused by sub-structures, such as a bar, that are not incorporated in the models used in this study. For stars younger than 10~Gyr, there is an excess around $L_\mathrm{z}=1850\sim2000\mathrm{~kpc~km~s^{-1}}$ and a deficit around $L_\mathrm{z}=2200\mathrm{~kpc~km~s^{-1}}$ that are not reproduced by the model. For stars older than 10~Gyr, there is an excess around $L_\mathrm{z}\sim1650\mathrm{~kpc~km~s^{-1}}$ and a deficit around $L_\mathrm{z}=1300\sim1500\mathrm{~kpc~km~s^{-1}}$.

\section{Verification of the dip in scale length}\label{sec:dipvalidation}
We test the robustness of the dip in scale length around $\tau = 10$~Gyr found in the fitting results for the observational data presented in Section~\ref{sec:result}.
Here, using the same method as in Appendix~\ref{sec:resultvalidation}, we compared the action distribution of 1,295 stars with ages between $\tau=9.5$ and 10.5~Gyr in the observational data and in two mock data sets generated from two different DF models. Model~1 follows the best-fit age dependence of each parameter obtained from the fitting result shown in Fig.~\ref{fig:result:spline}, same as the model used in Appendix~\ref{sec:resultvalidation}. Model~2 is identical to Model~1 except that it assumes a scale length $R_\mathrm{disc} = 1.72$~kpc, which is 3$\sigma$ larger than the \Rdisc value at the dip found in our results and similar to the scale length of the old thick disc (see Table.~\ref{tab:spline_result}).

\begin{figure*}	\includegraphics[width=\textwidth]{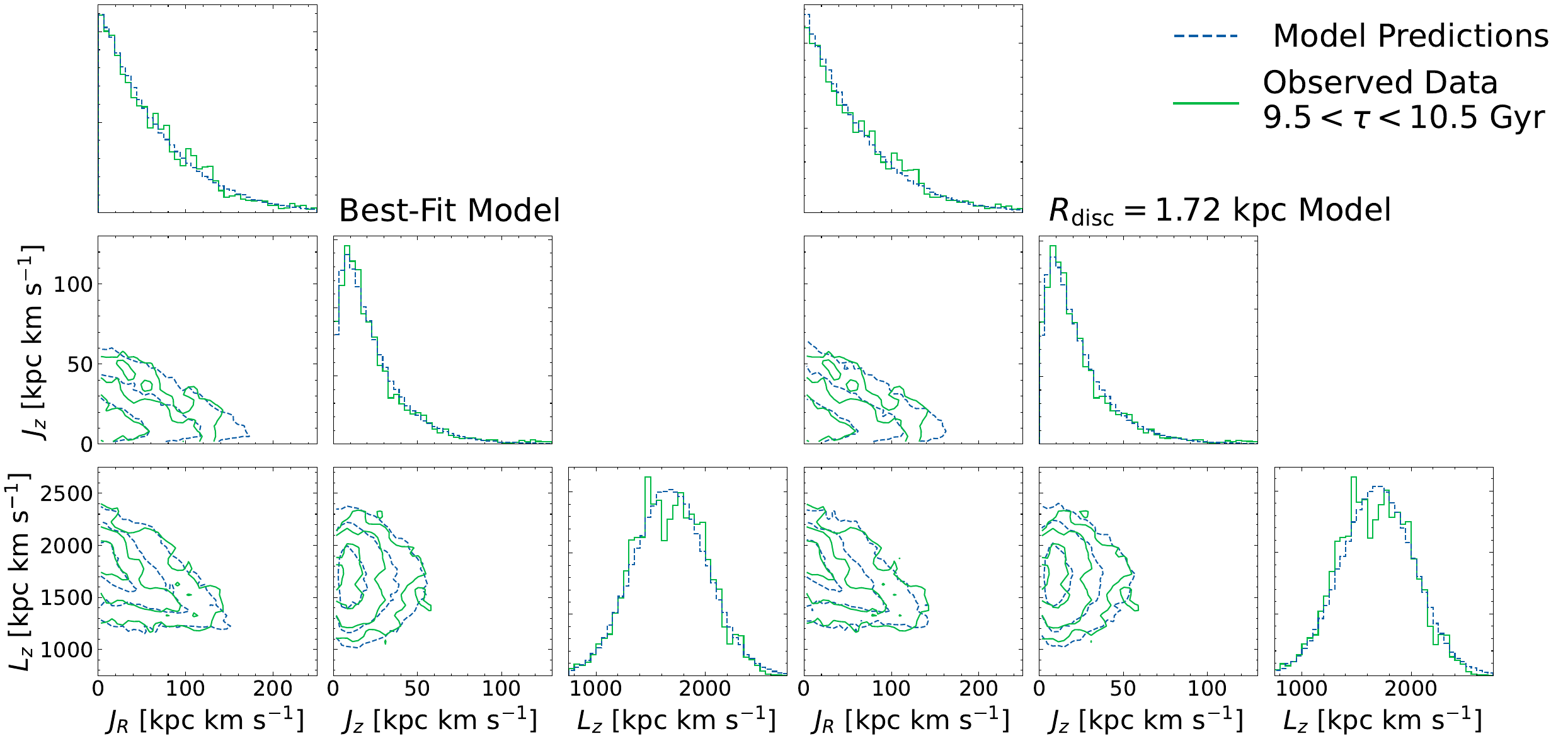}
    \caption{Comparison of action-space distributions for stars with ages in the range $9.5 < \tau < 10.5$~Gyr, corresponding to the epoch where a dip in $R_\mathrm{disc}$ is observed in Fig.~\ref{fig:result:spline}.
The left panel compares the observed data (green solid lines) with predictions from the best-fit model (blue dashed lines), using the parameter estimates shown in Fig.~\ref{fig:result:spline}.
The right panel shows a comparison with an alternative model in which $R_\mathrm{disc}$ is fixed at 1.72~kpc, simulating a scenario without the observed dip—while keeping all other parameters identical to those of the best-fit model. Both models appear to be in good agreement with the observed data, but when focusing on the shape of the distribution of the $J_R$ and $L_\mathrm{z}$ models, the best-fit model fits the observed data slightly better.
This comparison highlights the impact of the $R_{\mathrm{disc}}$ dip on the resulting action distributions.
}
    \label{fig:appendix:dip_actions}
\end{figure*}

 The left and right panels in Fig.~\ref{fig:appendix:dip_actions} compare the action distributions of the observational data with those of Model~1 and Model~2, respectively. The only difference between the models is a shift of approximately 0.2~kpc in $R_\mathrm{disc}$, yet both provide similarly good fits to the data.
 However, upon closer inspection, the $J_R$ histogram in Model~2 (right-hand panel) is slightly more strongly concentrated around zero than in the observational data or Model~1, indicating a mild mismatch with the observed distribution. There appears to be a subtle difference between the two models around $J_R=100\mathrm{~kpc~km~s^{-1}}$, where the data shows several fluctuations. Model~1 seems to go through the middle of this wobble, while Model~2 goes through the lower side of this wobble. $J_z$ shows little difference between the models. This may suggest that differences in \Rdisc do not have a significant effect on $J_z$. For $L_\mathrm{z}$, there is a bimodality around $L_\mathrm{z}\sim 1600 \mathrm{~kpc~km~s^{-1}}$ in the observation, but this is not reproduced in either model. This may be due to a resonance from the Galactic bar, which is not considered in our model. Still, Model~1 reproduces a distribution slightly closer to the observation than Model~2, as the sharpness around the peak is slightly weaker and has a wider distribution. 
 Hence, we consider that this dip in \Rdisc at $\tau\sim10$~Gyr is meaningfully inferred from the data.

\bsp	
\label{lastpage}
\end{document}